\documentclass[preprint,showpacs,preprintnumbers,amsmath,amssymb,12pt,a4paper,nofootinbib,pra]{revtex4}
\ifx\pdfoutput\undefined
\usepackage{graphicx,color}
\else
\usepackage[pdftex]{graphicx,color}
\fi

\usepackage{amsmath}
\usepackage{amssymb}
\usepackage{amsfonts}
\usepackage{amsthm}
\usepackage{exscale}
\usepackage{braket}
\usepackage{comment}
\usepackage{bm}

\newenvironment{matlabcode}[1]{}{}
\newcommand{\myeqnref}[1]{Eq.~(\ref{#1})}
\newcommand{\myeqnrefs}[1]{Eqs.~(\ref{#1})}

\newcommand{\field}[1]{\mathbb{#1}}

\newcommand{\figlength}{6in}

\newtheorem{theorem}{Theorem}

\newtheorem{corollary}{Corollary}
\newtheorem{claim}{Claim}

\theoremstyle{definition}
\newtheorem{definition}{Definition}

\begin{document}

\title{On the performance of two protocols: SARG04 and BB84}

\date{\today}

\author{Chi-Hang Fred Fung}
 \email{cffung@comm.utoronto.ca}
\author{Kiyoshi Tamaki}%
 \email{ktamaki@physics.utoronto.ca}

\author{Hoi-Kwong Lo}
 \email{hklo@comm.utoronto.ca}
\affiliation{%
Center for Quantum Information and Quantum Control,\\
Dept. of Electrical and Computer Engineering and Dept. of Physics,\\
University of Toronto, Toronto, Ontario M5S 3G4, Canada
}%

\begin{abstract}
We compare the performance of BB84 and SARG04, the later of which was proposed by V.~Scarani {\it et al.}, in Phys. Rev. Lett. {\bf 92}, 057901 (2004).
Specifically, in this paper, we investigate SARG04 with two-way classical communications and SARG04 with decoy states.
In the first part of the paper, we show that SARG04 with two-way communications can tolerate a higher bit error rate ($19.4\%$ for a one-photon source and $6.56\%$ for a two-photon source) than SARG04 with one-way communications ($10.95\%$ for a one-photon source and $2.71\%$ for a two-photon source).
Also, the upper bounds on the bit error rate for SARG04 with two-way communications are computed in a closed form by considering an individual attack based on a general measurement.
In the second part of the paper, we propose employing the idea of decoy states in SARG04 to obtain unconditional security
even when realistic devices are used.
We compare the performance of SARG04 with decoy states and BB84 with decoy states.
We find that the optimal mean-photon number for SARG04 is higher than that of BB84 when the bit error rate is small.
Also, we observe that SARG04 does not achieve a longer secure distance and a higher key generation rate than BB84, assuming a typical experimental parameter set.
\end{abstract}


\maketitle

\section{\label{sec:Introduction}Introduction}

Quantum key distribution (QKD) \nocite{Bennett1984}\cite{Ekert1991,Bennett1984} provides a way for two parties to expand a secure key that they initially share.
The best known QKD is the BB84 protocol published by Bennett and Brassard in 1984~\cite{Bennett1984}.
The BB84 protocol consists of two phases, the quantum transmission phase and the classical communication phase.
In the quantum phase, one of the two legitimate parties, Alice, sends quantum states to the other legitimate party, Bob.
The quantum states received by Bob are converted to classical bits by measurements.
In the classical communication phase, both parties discuss which bits to keep or discard.
They sacrifice some bits to test the error rate on the bit string.
If the error rate is too high, they abort the protocol.
For states that are retained, they perform bit error correction with the help of classical communications.
After that, Alice and Bob's bit strings are the same, but some information on them might have leaked to a potential eavesdropper, Eve.
To remove Eve's information, they apply privacy amplification to distill the final secret key.


The security of BB84 was not proved until many years after its introduction.
\nocite{Mayers2001}\nocite{Biham2000}\nocite{Lo1999}\nocite{Shor2000}
Among the proofs~\cite{Mayers2001,Biham2000,Lo1999,Shor2000}, the one by Shor and Preskill~\cite{Shor2000} is relevant to this paper.
Their simple proof essentially converts an entanglement distillation protocol (EDP)-based QKD proposed by Lo and Chau~\cite{Lo1999} to the BB84 protocol.
The EDP-based QKD has already been shown to be secure by~\cite{Lo1999} and the conversion successively leads to the security of BB84.

Security proofs of QKD protocols were further extended to explicitly accommodate the imperfection in practical devices~\cite{Gottesman2004,Inamori2005}.
One important imperfection is that the laser sources used in practice are coherent sources that occasionally emit more than one photon in each signal.
Thus, they are not single-photon sources that the other security proofs~\cite{Mayers2001,Biham2000,Shor2000} of BB84 assumed.
In particular, BB84 may become insecure when coherent sources with strong intensity are used.
For instance, Eve can launch an photon-number-splitting (PNS) attack, in which
 she blocks all single-photon pulses and splits multi-photon pulses.
She keeps one copy of each of the split pulses to herself and forwards another copy to Bob.
Although~\cite{Gottesman2004,Inamori2005} showed that secure QKD is still possible even with imperfect devices, the PNS attack puts severe limits on the distance and the key generation rate of unconditionally secure QKD.

A novel solution to the problem of imperfect devices in BB84 was proposed by Hwang~\cite{Hwang2003}, which uses extra test states--called the decoy states--to learn the properties of the channel and/or eavesdropping on the key-generating signal states.
Our group presented an unconditional security proof of decoy-state QKD~\cite{Lo2004,Lo2005}.
By combining the GLLP (Gottesman, Lo, L{\"{u}}kenhaus, and Preskill)~\cite{Gottesman2004} result with the decoy state idea~\cite{Hwang2003}, we showed that decoy state QKD can exhibit dramatic increase in distance and key generation rate compared to non-decoy protocols.
Moreover, our group proposed the idea of using the vacua or very weak coherent states as decoy states~\cite{Lo2004}.
Subsequently, practical protocols for QKD using a few decoy states were analyzed by Wang~\cite{Wang2005a,Wang2005b}, by our group~\cite{Ma2005b}, and by Harrington~\cite{Harrington2005}, thereby making the decoy idea more practical.
The first experimental implementation of a QKD using one decoy state was demonstrated by our group~\cite{Yi2005}.
Also, a decoy method using two-way classical communications was proposed by our group~\cite{Ma2005}.

Another attempt to combat PNS attacks was by Scarani {\it et al.}~\cite{Scarani2004} , who introduced a new protocol, called SARG04, which is very similar to the BB84 protocol.
The quantum state transmission phase and the measurement phase of SARG04 are the same as that of BB84, as both use the same four quantum states and the same experimental measurement.
The only difference between the two protocols is the classical post-processing phase.
Interestingly, with only a change in the post-processing phase, the protocol becomes secure even when Alice emits {\it two} photons, a situation under which BB84 is insecure.
This was proved by two of us~\cite{Tamaki2005}, who also proved the security of SARG04 with a single-photon source.
Specifically, we provided lower bounds of the bit error rate when one-way classical communications are used in the error correction and privacy amplification phases.
We also proposed a modified SARG04 protocol that uses the same six states as the original six-state protocol~\cite{Bruss1998,Lo2001}.
The security of SARG04 with a single-photon source was also proved by Branciard {\it et al.}~\cite{Branciard2005}.
They considered SARG04 implemented with single-photon sources and with realistic sources.
For the single-photon-source case, they provided upper and lower bounds of the bit error rate with one-way classical communications.
For the realistic-source case, they considered only incoherent attack by Eve and showed that SARG04 can achieve a higher secret key rate and a greater secure distance than BB84.
The SARG04 protocol was generalized by Koashi~\cite{Koashi2005} to the case of $N$ quantum states.
Another protocol that is similar to SARG04 is the B92 protocol~\cite{Bennett1992}, which uses two nonorthogonal quantum states.
The security of B92 with a single-photon source was proved by Tamaki {\it et al.}~\cite{Tamaki2003,Tamaki2004}.
On the other hand, Koashi~\cite{Koashi2004} proposed an implementation of B92 with strong phase-reference coherent light that was proved secure.

The fact that a modification to the classical communication part (from BB84 to SARG04) changes the foundation of security, i.e. making two-photon signals secure, is interesting.
Note that since the difference between BB84 and SARG04 is only in the classical data processing part, it is not difficult to perform SARG04 once the experiment of BB84 is available.
Thus, it is important to investigate the performance of SARG04 in order to determine which protocol one should perform.
This is our main motivation.

In this paper, we make an endeavour to study this interesting SARG04 protocol, but in different situations than that considered in~\cite{Tamaki2005,Koashi2005} and~\cite{Branciard2005}, and thus complementing their results.
Specifically, we provide upper and lower bounds of the bit error rate with {\em two-way classical communications} for single-photon sources and for two-photon sources.
Also, we consider implementations with realistic devices using decoy states with one-way classical communications.
Here, we allow {\em the most general attack} by Eve and study the key rate and distance properties of SARG04 in comparison with BB84.
Interestingly, under our most general attack assumption which was not considered in~\cite{Branciard2005}, we observe a different phenomenon than~\cite{Branciard2005}, that SARG04 has a lower key rate and a shorter secure distance than BB84.
However, our result shows that SARG04 is interestingly different from BB84 in one aspect in the realistic setting.
It is that the optimal mean photon number for SARG04 is higher than that for BB84, when the detector error probability is low.
This is because when the bit error rate gets smaller, the two-photon contribution to the key generation rate gets higher.

This paper makes use of two important existing techniques: QKD with two-way classical communications and the decoy-state method.
QKD with two-way communications in the error correction and privacy amplification phases was first proposed by Gottesman and Lo~\cite{Gottesman2003} as a method to achieve a higher tolerable bit error rate;
this method was later improved by Chau~\cite{Chau2002} to further increase the tolerable bit error rate of a six-state scheme.
The essence of QKD with two-way communications is that, by allowing Alice and Bob to communicate with each other, the qubits transmitted by Alice to Bob can be separated into two groups, one with a higher bit error rate than the other.
Thus, through two-way communications, they can discard the group with the higher bit error rate and retain the other group for further bit error correction and privacy amplification.
Intuitively, a QKD utilizing two-way communications should be superior to the case when only
one-way communications are used.
This was shown to be true for BB84 in~\cite{Gottesman2003}.
Here, we will show that this is also true for SARG04 for both single- and two-photon parts.
Especially for single-photon SARG04, we show that the lower bound with two-way communications is higher than the upper bound with one-way communications provided in~\cite{Branciard2005}.
When we analyze the security of SARG04 with realistic devices,
we will use the decoy-state method of~\cite{Lo2005} in order to achieve a long secure distance.

\begin{table}
\caption{\label{table-Summary}Summary of results for SARG04.  The numbers marked with $^\dagger$ are the results of this paper.
The bounds on the secure distance are specific for the experimental parameters from the Gobby-Yuan-Shields (GYS) experiment~\cite{Gobby2004}.
}
\begin{ruledtabular}
\begin{tabular}{lcc}
\multicolumn{3}{c}{Bit error rate of SARG04 with single-photon source} \\
& one-way & two-way \\
Upper bound & 14.9\%~\cite{Branciard2005} & $1/3^\dagger$ \\
Lower bound & 9.68\%~\cite{Tamaki2005,Branciard2005} and 10.95\% (with preprocessing)~\cite{Branciard2005} & $19.9\%^\dagger$ \\
\hline
\multicolumn{3}{c}{Bit error rate of SARG04 with two-photon source} \\
& one-way & two-way \\
Upper bound & N/A & $22.56\%^\dagger$ \\
Lower bound & 2.71\%~\cite{Tamaki2005} & $6.56\%^\dagger$ \\
\hline
\multicolumn{3}{c}{Secure distance using decoy states with realistic source} \\
& BB84 & SARG04 \\
Upper bound & 207.7 km~\cite{Lo2005} & $207.7 \text{ km}^\dagger$ \\
Lower bound & 141.8 km~\cite{Lo2005} & $97.2 \text{ km}^\dagger$ \\
\end{tabular}
\end{ruledtabular}
\end{table}
We have tabulated the results of this paper on bounds of bit error rate and secure distance, along with known results, in Table.\ref{table-Summary}.
The six numbers on the right column are results of this paper, while existing results are cited on the left column.
The bounds on the secure distance listed are specific for the experimental parameters from the Gobby-Yuan-Shields (GYS) experiment~\cite{Gobby2004}.

The organization of the paper is as follows:
We first review some existing techniques for the security proof in Section~\ref{sec:Preliminaries}, which provide a basis for the development of the results of this paper. In Section~\ref{sec-assumptions}, we summarize the assumptions we make in this paper. In Section~\ref{sec:one-two-photon-source}, we develop a SARG04 protocol with two-way classical communications with one- and two-photon sources.
In Section~\ref{sec-SARG04-deocy}, we consider SARG04 in a realistic setting, where imperfect laser sources and detectors are used.
Finally, concluding remarks are provided in Section~\ref{sec:conclusion}.

We note that an independent work on SARG04 with decoy states was also studied in~\cite{Li2005}.

\section{\label{sec:Preliminaries}Preliminaries}

In this section,
we review some bases for the security proof in this paper.
First, we briefly review an entanglement distillation protocol (EDP) and its relation with the security of QKD, where we especially review the security proof of BB84 by Shor and Preskill~\cite{Shor2000}.
Secondly, we explain how SARG04 works, and we construct an EDP protocol that is equivalent to SARG04 protocol.
We furthermore mention the property of the density matrix in the EDP protocol for the later convenience.
Thirdly, we explain the key generation rate for BB84 and SARG04, assuming realistic devices and one-way classical communications.
Next, we describe the decoy method in BB84 and SARG04.
Finally, we review QKD with two-way classical communications.


\subsection{EDP and its relation with QKD}

\subsubsection{EDP}

The goal of an entanglement distillation protocol (EDP) is to distill nearly perfect EPR pairs from noisy EPR pairs initially shared between two distant parties, Alice and Bob.
Any bipartite density matrix describing Alice and Bob's qubit system, $\rho$, can be expressed in the Bell basis, which is composed of the four orthogonal Bell states:
\begin{equation}
\label{eqn-Bellstates}
\begin{split}
\ket{\Phi^{\pm}} &= ( \ket{00} \pm \ket{11} ) /\sqrt{2} \\
\ket{\Psi^{\pm}} &= ( \ket{01} \pm \ket{10} ) /\sqrt{2} .
\end{split}
\end{equation}
Taking $\ket{\Phi^+}$ as the reference state, the diagonal of $\rho$ in the Bell basis
\begin{equation}
\label{eqn-Bellerrors}
\begin{split}
p_I &\triangleq \bra{\Phi^+}\rho\ket{\Phi^+} \\
p_X &\triangleq \bra{\Phi^-}\rho\ket{\Phi^-} \\
p_Z &\triangleq \bra{\Psi^+}\rho\ket{\Psi^+} \\
p_Y &\triangleq \bra{\Psi^-}\rho\ket{\Psi^-}
\end{split}
\end{equation}
represent the probabilities of applying, respectively, the Pauli $I$, $X$, $Z$, and $Y$ operators to the either one of the qubits of the bipartite system.
In the view of an EDP,
a pool of $\ket{\Phi^+}_{AB}$ state is prepared by Alice.
She keeps system $A$ of every pair and sends system $B$ of every pair to Bob.
Due to the presence of noise in the quantum channel,
system $B$ may undergo bit and/or phase flip errors and the probabilities of the various types of errors are represented by $p_I$ (no error), $p_X$ (bit flip error), $p_Z$ (phase flip error), and $p_Y$ (bit and phase flip error).
In the paper by Bennett, DiVincenzo, Smolin, and Wootters (BDSW)~\cite{BDSW1996}, they assume that all of the pairs are described by the same density matrix,
and the job of an EDP is to correct the errors using only local operations and classical communications (LOCCs), leaving Alice and Bob with a pool of $\ket{\Phi^+}_{AB}$ states.
Several methods of EDP's were proposed in BDSW~\cite{BDSW1996} including the hashing method and the recurrence method.
Many of these methods assume that the initial density matrix $\rho$ is Bell-diagonal.

\subsubsection{EDP-based QKD protocol}

EDP's are closely related to QKD protocols.
The connection between them is that
if Alice and Bob share almost perfect EPR pairs that are pure, then the pairs are almost unentangled with Eve's system.
Thus, the information leaked to Eve is negligible, and they can obtain an unconditionally secure key by measuring the EPR pairs.
Thus, the purpose of a QKD protocol can be viewed as a procedure for Alice and Bob to share almost perfect EPR pairs, which is the purpose of an EDP.
In order to run an EDP, they need to know the error rates on the noisy EPR pairs and the job of the error rate estimation is the first part of a QKD protocol.
After the error rates are upper bounded, the second part of the QKD involves running an EDP to distill almost perfect EPR pairs.
In essence, a QKD can be regarded as consisting of an error rate estimation part and an EDP part.
Note that the eavesdropping attack by Eve who has read/write access to the quantum channel appears to Alice and Bob as noise of the channel.

\begin{figure}
\centering
\includegraphics[width=\figlength]{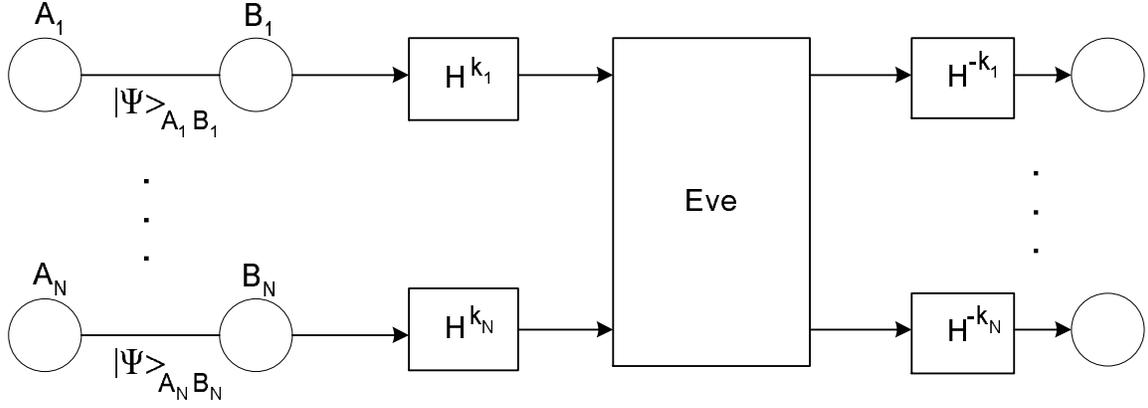}
\caption{\label{fig-bb84epr1}
An EDP version of the BB84 protocol.
Shor and Preskill~\cite{Shor2000} showed that it can be reduced to BB84.
Note that only the EPR pairs to which Alice and Bob apply the same rotations are shown;
EPR pairs with different rotations are discarded.
}
\end{figure}

An EDP-based QKD using quantum computers was proposed by~\cite{Lo1999} and a modified version of it~\cite{Shor2000} (shown in Fig.~\ref{fig-bb84epr1}) is as follows:
Alice prepares $N$ EPR pairs $\ket{\Psi}_{A_i B_i}=(\ket{0_z}_{A_i} \ket{0_z}_{B_i} + \ket{1_z}_{A_i} \ket{1_z}_{B_1})/\sqrt{2}, \text{ for } i \in [1,N]$.
She randomly chooses whether to apply a Hadamard gate $H$ on system $B$ (i.e. $k_i=0,1$) before sending it to Bob through Eve.
Eve may perform the most general attack on all Bob's qubits.
Bob randomly chooses whether to apply the Hadamard.
They discard the EPR pairs to which Alice and Bob apply different operations.
Alice and Bob choose some of the EPR pairs as test qubits.
They measure the test qubits in the $Z$ basis and compare the measurement results publicly to estimate the bit error rate of the test qubits.
The random sampling theorem then asserts that the rest of the untested qubits (code bits)
have asymptotically the same bit error rates as the test bits with high probability.
Since the bit errors and the phase errors are symmetrized by the random Hadamard gate on Bob's qubits,
the phase error rate on code bits is asymptotically equal to the bit error rate on code bits, i.e. for BB84
\begin{eqnarray}
\label{eqn-BB84epeb}
e_p&=&e_b.
\end{eqnarray}
Once Alice and Bob know the good estimates the error rates, they can each obtain the bit and phase error syndromes using quantum computers.
Alice then sends her syndromes to Bob who will then correct his qubits by applying $Z$ and $X$ operations so that his syndromes match Alice's syndromes.
After the successful distillation, they now share EPR pairs that have high fidelity with the pure state $\ket{\Phi^+}_{AB}^{\otimes M}$ (where $M$ is the number of the EPR pairs Alice and Bob share).
They each measure their halves of the pair in the $Z$ basis to produce a common secure key on which Eve has negligible information.

We can associate the four probabilities $p_I, p_X, p_Z, p_Y$ with two (dependent) binary random variables, $X$ and $Z$, which represent the bit and phase errors, respectively.
With this notation, the uncertainty in the bit flip error is $H(X)=H_2(p_X+p_Y)$ and in the phase flip error is $H(Z)=H_2(p_Z+p_Y)$,
where $H_2(p)=-p \log_2(p) - (1-p) \log_2(1-p)$ is the binary entropy function.
The mutual information between the bit and phase errors is $I(X;Z)=H(X)-H(X|Z)$.

The key generation rate of the EDP-based QKD using one-way classical communications is~\cite{BDSW1996}
\begin{eqnarray}
R
&=& 1- H(X,Z) \\
\label{eqn-rateoneway}
&=& 1- H(X) - H(Z|X) .
\end{eqnarray}
The second term in the last equation is concerned with the number of rounds of random hashing for determining the bit error patterns,
and
the third term is concerned with the number of rounds of random hashing for determining the phase error patterns given that the bit error patterns are known.
One drawback with the EDP-based QKD is that it requires the preparation of EPR pairs and the use of quantum memory and computers, which are challenging to implement in practice in the near future.
Thus, it is more desirable to use prepare-and-measure QKD protocols, in which Alice only needs to prepare qubits and send them to Bob, and Bob only needs to measure them immediately after receiving them;
no quantum memory and quantum computers are needed.


\subsubsection{BB84 protocol}

In Shor and Preskill's proof~\cite{Shor2000}, they showed that the EDP-based QKD can be reduced to BB84, a prepare-and-measure protocol that does not require the use of quantum computers.
Their proof relies on the use of CSS codes to decouple the bit error correction and the phase error correction.
They showed that phase error correction is not necessary; as long as phase error correction {\it could have been performed}, the protocol is secure.
Thus, the phase error correction step with quantum decoding is replaced by a privacy amplification step where classical bits of the raw key are XOR'ed to form the final key.
Since the phase error correction step is removed, Bob's final $Z$ measurement in the EDP-based QKD can be moved to before the bit error correction step.
Here, note that all of the hashing for the bit error correction is in the $Z$ basis, which commutes with Bob's final $Z$ measurements.
Only one-way communications are needed in the bit error correction step in Shor-Preskill's proof.
This is because Alice and Bob both compute the bit error syndromes but only Alice sends her syndromes to Bob.
Bob then applies the appropriate bit-flip operations on his bit string so as to match his syndromes with Alice's syndromes.
Using \myeqnref{eqn-rateoneway}, the key generation rate of the BB84 protocol resulting from the use of CSS codes is
\begin{eqnarray}
R
&=& 1- H(X) - (H(Z)-I(Z;X)) \\
\label{eqn-rategeneral1}
&=& 1- H_2(e_b) - H_2(e_p) + I(Z;X)
\end{eqnarray}
where $e_b=p_X+p_Y$ is the bit error rate
 and $e_p=p_Z+p_Y$ is the phase error rate.
The bit error rate $e_b$ is estimated in BB84 through public communications between Alice and Bob.
It is important to note that the phase error rate $e_p$ can be estimated from $e_b$ using \myeqnref{eqn-BB84epeb}.
The mutual information term in \myeqnref{eqn-rategeneral1} can be determined by $p_X$, $p_Y$, $p_Z$.
However, only $e_b=p_X+p_Y$ and $e_p=p_Z+p_Y$ are known and $p_Y$ is not known.
Thus, we consider the worst-case value of $p_Y$ (which corresponds to having no mutual information between bit and phase errors) to find the worst-case value of the key generation rate.
In the worst-case scenario, the highest tolerable bit error rate can be found by solving $1=2H_2(e_b)$.
This gives $e_b=11.0\%$~\cite{Shor2000}, at which the key generation rate is zero.

\subsection{The SARG04 protocol\label{sec-SARG04}}

In this paper, we consider the SARG04 protocol~\cite{Scarani2004}, which is a prepare-and-measure protocol.
In fact, the quantum phase of SARG04 is the same as that of BB84;
so it can easily be seen that SARG04 is a prepare-and-measure protocol as BB84 is.

Let us explain how SARG04 works.
In SARG04 there are four quantum states, $\ket{\varphi_i}, i=0,\ldots,3$:
\begin{equation}
\begin{split}
\ket{\varphi_0}&=\beta \ket{0_x} + \alpha \ket{1_x} \\
\ket{\varphi_m}&=R^{-m} \ket{\varphi_0}, \: m=0,\ldots,3,
\end{split}
\end{equation}
where $\alpha \equiv \sin(\pi/8)$, $\beta \equiv \cos(\pi/8)$, and
$R \equiv \cos(\pi/4) I + \sin(\pi/4) (\ket{1_x}\bra{0_x}-\ket{0_x}\bra{1_x})$ is a $\pi/2$ rotation around the $Y$ basis.
Note that $\ket{\varphi_0}$ and $\ket{\varphi_2}$ are orthonormal, and thus form a basis.
The same can be said for $\ket{\varphi_1}$ and $\ket{\varphi_3}$.
The four states are divided into four sets, $\{R^K \ket{\varphi_0},R^K \ket{\varphi_1}\}, K\in[0,3]$,  in which one represents logic $0$ and the other logic $1$.
The steps for the SARG04 protocol with a $\nu$-photon source ($\nu=1,2$) and one-way communications are as follows:
\begin{enumerate}
\item Alice sends a sequence of $N$ signals to Bob.
For each signal, Alice randomly chooses one of the four sets and sends one of the two states in the set to Bob.
\item For each signal, Bob performs the polarization measurement using one of the two bases randomly. If his detector fails to click, then he broadcasts this fact, and Alice and Bob discard all the corresponding data.
\item For each signal, Alice publicly announces the choice of the set from which the state was selected.
\item For each signal, Bob compares his measurement outcome to the two states in the set.
If his measurement outcome is orthogonal to one of the states in the set, then he concludes that the other state has been sent, which is a conclusive result.
On the other hand, if his measurement outcome is not orthogonal to either of the states in the set, he concludes that it is an inconclusive result.
He broadcasts if he got the conclusive result or not for each signal.
\item Alice randomly chooses some bits as test bits and announces their locations.
Bob estimates the bit error rate $e_{\nu}$ from the test bits by taking the ratio of the number of incorrect conclusive test bits to the total number of conclusive test bits.
If $e_{\nu}$ is too high, they abort the protocol.
\item Alice and Bob retain only the conclusive untested bits.
\item They perform bit error correction and privacy amplification on the remaining bit string.
\end{enumerate}

\begin{figure}
\centering
\includegraphics[width=\figlength]{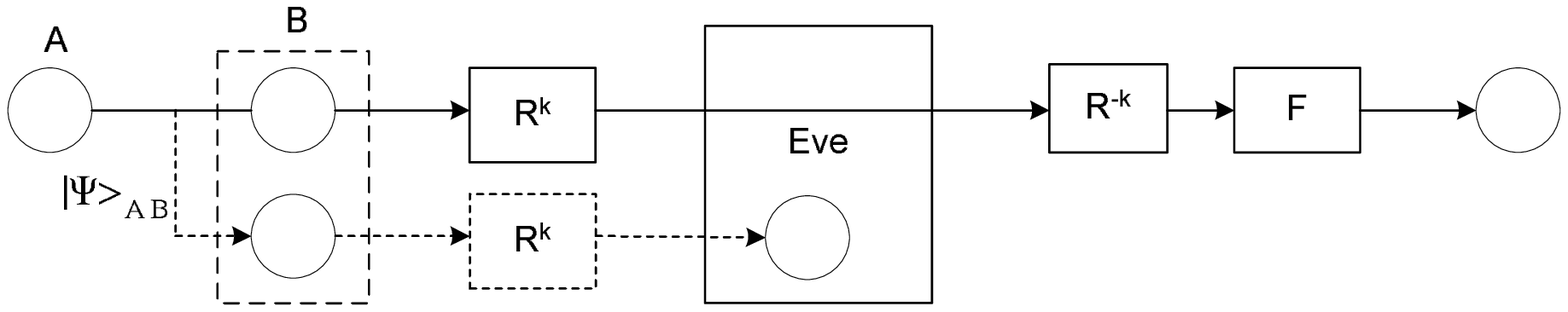}
\caption{\label{fig-sarg04epr1}
An EDP version of the SARG04 protocol.
Alice prepares an entangled states $\ket{\Psi}_{AB}=(\ket{0_z}_A \ket{\varphi_0^{\otimes \nu}}_B + \ket{1_z}_A \ket{\varphi_1^{\otimes \nu}}_B)/\sqrt{2}$ where $\nu=1,2$ corresponds to the number of photons emitted by Alice.
She applies a random rotation on system $B$ before sending it to Bob through Eve.
In the case of $\nu=2$, Eve retains one qubit of system $B$ and sends the other to Bob.
Although, only one entangled state is shown for simplicify, one should be reminded that Eve may perform the most general attack on the $N$ entangled states as in Fig~.\ref{fig-bb84epr1}.
}
\end{figure}
We construct an EDP version of the SARG04 protocol, which is shown in Fig.~\ref{fig-sarg04epr1}.
The EDP version lends itself to an easy extension with two-way classical communications and also a simplified analysis on the bounds on the bit error rates, both of which will be studied in detail later in this paper.
We consider Alice having a $\nu$-photon source, $\nu=1,2$.
For each signal, she first prepares an entangled state $(\ket{0_z}_A \ket{\varphi_0^{\otimes \nu}}_B + \ket{1_z}_A \ket{\varphi_1^{\otimes \nu}}_B)/\sqrt{2}$ and randomly applies a rotation $(R^K)^{\otimes \nu}$ to system B which is then sent to Bob through Eve.
Eve applies the most general attack on all the $N$ signals jointly.
We assume that Eve always sends a qubit state or a vacuum state to Bob, which is related to
the assumption we describe in Section~\ref{sec-assumptions}.
Bob, upon receiving the qubit, performs the inverse rotation $R^{-K'}$ and a filtering operation whose successful operation is described by the Kraus operator as $F=\sin\frac{\pi}{8} \ket{0_x}_B \bra{0_x} + \cos\frac{\pi}{8} \ket{1_x}_B \bra{1_x}$. Here, the successful filtering corresponds to a conclusive result~\cite{Tamaki2003,Tamaki2004} in the prepare-and-measure SARG04 protocol.
Alice and Bob then publicly exchange $K$ and $K'$ and keep the pairs with $K=K'$.
They randomly choose some states (test bits) and perform $Z$ measurements on the states. Then, they compare their measurement outcome publicly in order to estimate the bit error rate on the remaining pairs (code bits). This gives us a good estimation of the bit error rate on code bits thanks to the
random sampling theorem.
On the other hand, the phase error rate on the code bits is estimated from the bit error rate on the code bits by the theorem below. After the estimation, they choose a CSS code that is sufficient to correct all the bit and phase errors.
After the error correction, they share maximally entangled states from which they
perform $Z$ measurements to obtain a secure key.
It is important to note that the phase error rate of the code bits can be estimated from the bit error rate. Thanks to this estimation, Alice and Bob
do not need to perform test bit in $X$ measurement, thus we can equivalently convert our EDP protocol to the prepare-and-measure protocol by the Shor-Preskill's arguments.








\begin{theorem}[Density matrix of one-photon SARG04]
For the one-photon case,
the diagonal elements of the density matrix of the EPR pair shared between Alice and Bob in the Bell basis is
\begin{eqnarray}
\label{eqn-single-densitymatrix}
p_X&=&e_b -a \nonumber \\
p_Z &=&\frac{3}{2} e_b - a \\
p_Y &=& a, \nonumber
\end{eqnarray}
where $e_b$ is the bit error rate, and $e_b/2 \leq a \leq e_b$.
\end{theorem}
\begin{proof}
See Appendix~\ref{app-SARG04-densitymatrix}.
\end{proof}
There are two differences between this density matrix and that for BB84:
(i) there is a factor of $\frac{3}{2}$ in $p_Z$ (whereas the factor is one in BB84),
and (ii) $a$ is no smaller than $e_b/2$ (whereas $a$ can be as small as zero in BB84).
Such a restriction in $a$ gives rise to mutual information between bit and phase errors (see also~\cite{Tamaki2005}).
This is because for bit and phase errors to be independent (i.e. no mutual information), $p_Y = a$ has be to equal to $\frac{3}{2}e_b^2$.
But, this is outside the range $e_b/2 \leq a \leq e_b$
for $e_b < \frac{1}{3}$ which is the case of interest.
The lower bound on the bit error rate for the one-photon case can be found by solving
$0=1-H_2(e_b)-H_2(3e_b/2)+I(X;Z)$, which gives $e_b=9.68\%$~\cite{Tamaki2005,Branciard2005}.
Note that~\cite{Branciard2005} provided a better bound of $e_b=10.95\%$ with data preprocessing.

\begin{theorem}[Density matrix of two-photon SARG04]
\label{thm-SARG04-densitymatrix-two}
For the two-photon case,
the diagonal elements of the worst case density matrix is
\begin{eqnarray}
\label{eqn-two-densitymatrix}
p_X&=&e_b-a \nonumber \\
\label{eqn-two-1}
p_Z &\leq& x e_b + g(x) -a , \forall x \\
p_Y&=&a, \nonumber
\end{eqnarray}
where $g(x)=\frac{1}{6} (3-2x+\sqrt{6-6\sqrt{2}x+4x^2})$ and $0 \leq a \leq e_b$.
\end{theorem}
\begin{proof}
See Appendix~\ref{app-SARG04-densitymatrix}.
\end{proof}
In this case, $a$ is allowed to be zero.
Thus, the lower bound on the bit error rate for the two-photon case can be found by
minimizing \myeqnref{eqn-rategeneral1} over $a$, which leads to having no mutual information between bit and phase errors (i.e. $I(X;Z)=0$).
Solving $0=1-H_2(e_b)-H_2(\min_{x} x e_b+g(x))$ gives $e_b=2.71\%$~\cite{Tamaki2005}.

\subsection{Privacy amplification for multi-photon signals}

In real-life implementation, a weak laser pulse is often used to simulate a single-photon source.
However, since it actually emits weak coherent states, the laser outputs contain some multi-photon states in addition to the desired single-photon states.
The phases of the coherent pulses are assumed to be randomized in a
traditional laser source.
Because of this, the coherent states of the laser output reduce to classical mixtures of photon-number states with a Poisson distribution.
One important idea from GLLP~\cite{Gottesman2004} is that the amount of privacy amplification needed when multi-photon signals are present is the same as if only the key-generating signals are present.
To illustrate the idea, let us consider the key-generation rate for BB84.
For BB84, the final key can only be generated by using the single-photon states.
If Alice and Bob knew the locations of the single-photon states, they could discard all other multi-photon states and apply error correction and privacy amplification only to the single-photon states.
In this case, they could achieve a rate of
\begin{eqnarray}
R_{\text{\tiny BB84}} &=& -Q_1 f(e_1) H_2 (e_1) + Q_1 [1-H_2(e_1)],
\end{eqnarray}
where $e_n$ is the bit error rate of the $n$-photon signal states,
$Q_n$ is the gain\footnote{The gain of a particular type is the probability that the transmitted signal of that type is sent by Alice and Bob gets a conclusive result.} of the n-photon signal state, and
$f(x)$ is the error correction efficiency as a function of error rate.
The first term is concerned with number of rounds of random hashing for determining the bit error patterns and the $H_2(e_1)$ in the second term is concerned with the privacy amplification.
Note that the bit error rate $e_1$ is used for the privacy amplification term because of \myeqnref{eqn-BB84epeb}.
For BB84, Bob's result is conclusive when Bob obtains bit value by the 
same measurement basis as the one that Alice has chosen.

Note that the above rate is achieved only when Alice and Bob know the locations of the single-photon states, which is not the case that Bob uses a threshold detector.
One method to achieve unconditional security without Alice and Bob knowing the locations of the single-photon states was
proposed by~\cite{Gottesman2004}.
The idea is that privacy amplification applied to all bit string is equivalent to that applied only to the bit string stemmed from the single-photon states as if the locations of them are known.
To show this, we consider the bit value produced by $k_1 \cdot V_1 \oplus k_M \cdot V_M$, 
where $k_1$ and $k_M$ are the bit string stemmed from the single- and multi-photon states after bit error correction, and $V_1$ and $V_M$ are random strings in a hash function having the same lengths as $k_1$ and $k_M$ respectively. The first term of $k_1 \cdot V_1 \oplus k_M \cdot V_M$ corresponds to privacy amplification applied to single-photon states only, while the second term is some bit (possibly known to Eve).
Since the first term is private to Alice and Bob, even if the second term is completely known to Eve, the sum is still private to Alice and Bob.
With this idea, the key generation rate can be improved by considering privacy amplification applied only to single-photon states:
\begin{eqnarray}
\label{eqn-decoyBB84rate1}
R_{\text{\tiny BB84}} &=& -Q_\mu f(E_\mu) H_2 (E_\mu) + Q_1 [1-H_2(e_1)].
\end{eqnarray}
In this paper, we consider SARG04 which is secure with single-photon and two-photon states.
In this case, the key generation rate is~\cite{Tamaki2005}
\begin{eqnarray}
\label{eqn-decoySARG04rate1}
R_{\text{\tiny SARG04}} &=& -Q_\mu f(E_\mu) H_2 (E_\mu) + Q_1 [1-H(Z_1|X_1)] + Q_2 [1-H(Z_2|X_2)],
\end{eqnarray}
where the $Z_n$ ($X_n$) is a random variable corresponding to the phase (bit) error for the n-photon state.
The first term is the fraction of EPR pairs spent for error correction, the second term is the contribution to the key rate from the single-photon states, and the third term is the contribution from the two-photon states.
Note that the mutual information between the bit and phase errors is included.
According to Theorem~\ref{thm-SARG04-densitymatrix-two}, the mutual information between $X_2$ and $Z_2$ can be zero, meaning $H(Z_2|X_2)=H(Z_2)$.

In \myeqnref{eqn-decoyBB84rate1} and \myeqnref{eqn-decoySARG04rate1}, the overall gain $Q_\mu$ and the overall bit error rate $E_\mu$ are parameters that Alice and Bob can estimate through public communications.
On the other hand, the gain $Q_1$ (and $Q_2$ for SARG04), and the bit error rate for the single-photon states $e_1$ (and $e_2$ for SARG04) cannot be directly estimated.
One way to estimate $e_1$ and $Q_1$ (and $e_2$ and $Q_2$) is to consider the worst situation for Alice and Bob.
For instance, in BB84, we can pessimistically assume that all the errors happen only in the single-photon detection events, leading to
$e_1=E_\mu Q_\mu/Q_1$ and $Q_1=Q_\mu-p_{multi}/2$, where $p_{multi}$ is the probability of Alice emitting multiple-photon states (see~\cite{Lo2005}).
However, this gives a low key generation rate and a short secure distance.
Another way to
estimate $e_n$ and $Q_n$ is to use the decoy-state method in~\cite{Lo2005}, which we explain next.
Using this method, the key generation rate and the secure distance can be greatly increased.

\subsection{Decoy-state method}

In the security analysis with decoy states,
we assume using the infinite-decoy-state method of~\cite{Lo2005} for the simplicity of analyses.
Let us first define the yield $Y_n$, the bit error rate $e_n$, the gain $Q_n$.
The yield, $Y_n$, is defined as the probability that
Bob's measurement outcome is conclusive conditional on Alice's $n$-photon emission:
\begin{eqnarray}
\label{eqn-def-yield}
Y_n &\triangleq&
Pr\{ \text{Bob's result is conclusive} | \text{Alice sent $n$-photon state} \} .
\end{eqnarray}
The yield is basically a sum of the probabilities of the error events and the error-free events.
The fraction of the error-event probability is the bit error rate $e_n$:
\begin{eqnarray}
\label{eqn-def-ber}
e_n &\triangleq&
Pr\{ \text{Bob's result is incorrect} | \nonumber \\
&& \phantom{Pr\{} \text{Bob's result is conclusive} \wedge \text{Alice sent $n$-photon state} \} .
\end{eqnarray}
The gain of the $n$-photon state is
\begin{eqnarray}
\label{eqn-def-Qn}
Q_n &\triangleq&
Pr\{ \text{Bob's result is conclusive} \wedge \text{Alice sent $n$-photon state} \} \\
&=& Y_n e^{-\mu} \mu ^n / n! .
\end{eqnarray}
The key of the decoy method is to consider the two equations for the overall gain $Q_\mu$ and the overall bit error rate $E_\mu$.
The overall gain is the weighted average of the yields of all n-photon states:
\begin{eqnarray}
\label{eqn-def-Qmu}
Q_\mu &=& Y_0 e^{-\mu} + Y_1 e^{-\mu} \mu + \cdots + Y_n e^{-\mu} (\mu^n/n!) + \cdots .
\end{eqnarray}
The overall QBER is the weight average of the errors of all n-photon states:
\begin{eqnarray}
\label{eqn-def-Emu}
E_\mu &=& \frac{1}{Q_\mu}  \sum_{n=0}^{\infty}  Y_n e^{-\mu} (\mu^n/n!) e_n
\end{eqnarray}
The main point of the method is to vary the laser intensity $\mu$ over all non-negative values randomly.
Each value of $\mu$ is associated with one equation for $Q_\mu$ and one for $E_\mu$.
Thus, by varying $\mu$, we have a set of linear equations of $Y_n$ and $e_n$, which can then be solved.
The states that are used for the determination of $Y_n$ and $e_n$ with the different $\mu$'s are the decoy states, which will not be used to generate the final key.
Another set of states, the signal states, will be used for key generation and are outputs from one laser intensity only.
To make sure that $Y_n$ and $e_n$ estimated from the decoy states are good estimates of $Y_n$ and $e_n$ for the signal states, we randomize the locations of both states so that Eve can only act equally on them.
Once we have good estimates of $Y_n$ (thus, $Q_n$) and $e_n$, we can determine the achievable key-generation rate by using \myeqnref{eqn-decoyBB84rate1} for BB84 and \myeqnref{eqn-decoySARG04rate1} for SARG04.
For SARG04, we use the relations between the phase and bit error rates in \myeqnref{eqn-single-densitymatrix} and \myeqnref{eqn-two-densitymatrix} to determine the phase error rates from the bit error rates.

For BB84, the expected values for the yields and the bit error rates without any eavesdropping are~\cite{Lo2005}
\begin{eqnarray}
\label{eqn-BB84Yn}
Y_{n,\text{\tiny BB84}}&=&[\eta_n + (1-\eta_n) p_{dark}]/2 \\
\label{eqn-BB84en}
e_{n,\text{\tiny BB84}}&=&(\eta_n \frac{e_{detector}}{2} + (1-\eta_n) p_{dark} \frac{1}{4})/Y_{n,\text{\tiny BB84}},
\end{eqnarray}
where $\eta_n$, $p_{dark}$, and $e_{detector}$ are transmission efficiency
for an $n$-photon signal, the probability that the detector clicks when the input is a vacuum state, and a parameter representing the misalignment in the detector, respectively.
The presence of any eavesdropping would deviate the actual values of them and thus would be caught by Alice and Bob.
For SARG04, we will derive similar formulas for $Y_n$ and $e_n$ later in this paper, and also we will describe
the SARG04 protocol with decoy states.

\subsection{QKD with two-way classical communications\label{sec-prelim-two-way}}

In Shor-Preskill's proof, they showed that applying the bit and phase error corrections with CSS code followed by $Z$ measurements to a pool of noisy EPR pairs is equivalent to applying the $Z$ measurement followed by bit error correction and privacy amplification.
This order swapping is applicable to any pool of noisy EPR pairs characterized by some $(p_X,p_Y,p_Z)$.
Imagine that, before the bit and phase error corrections and the final $Z$ measurements , we insert an extra operation on the EPR pairs that changes the pairs to have some other characteristics $(p_X',p_Y',p_Z')$.
One reason that we want to insert such an extra operation is to increase the highest tolerable bit error rate of a QKD protocol.
Since, after this extra operation, we are also left with a pool of noisy EPR pairs, we can invoke the Shor-Preskill's argument to move the final $Z$ measurements to before the bit and phase error correction steps.
However, this is not (yet) a prepare-and-measure protocol since Shor-Preskill's proof only brings the $Z$ measurements to after the extra operation.
If this extra operation commutes with the $Z$ measurements, then we can swap their order and turn it into a prepare-and-measure protocol.

A specific operation for this extra operation was considered by Gottesman and Lo~\cite{Gottesman2003}.
Their operation commutes with the $Z$ measurements (so is compatible with prepare-and-measure protocols) and is composed of a sequence of steps applied to the EPR pairs.
There are two types of steps, a B step and a P step.
As the names imply, a B step (P step) is meant to improve the bit (phase) error rate of the EPR pairs.
A B step requires two-way classical communications for exchanging information between Alice and Bob.
Hence, prepare-and-measure protocols derived from using this technique requires two-way classical communications.

\begin{figure}
\centering
\includegraphics{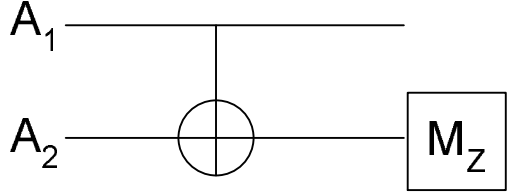}
\caption{\label{fig-bstep}
B step: Alice performs a bilateral XOR on her halves of two EPR pairs (the circuit of which is depicted) and Bob performs the same on his halves of the EPR pairs.
They measure the target qubits in the $Z$ basis.
If their measurement results are same, they keep the source EPR pair and discard the target;
otherwise, they discard both EPR pairs.
}
\end{figure}

\begin{definition}[B step~\cite{Gottesman2003}]
A B step, shown in Fig.~\ref{fig-bstep}, consists of
Alice and Bob together performing a bilateral XOR on two EPR pairs randomly chosen and comparing their $Z$ measurement results of the target pair.
If their results are the same, they keep the source EPR pair and discard the target EPR pair.
If they are different, they discard both pairs.
When the two EPR pairs initially have no bit error or both have a bit error, the measurement results will be the same.
When only one of the two pairs has a bit error, the measurement results will be different.
The bilateral XOR is equivalent to two measurements of $Z \otimes Z $, one by Alice and one by Bob.
Thus, a B step commutes with the final $Z$ measurements in a prepare-and-measure protocol.
Suppose that initially the
EPR pairs are in the state $(p_X,p_Y,p_Z)$, applying a B step to every pair of EPR pairs leads to a smaller set of surviving pairs with a new state $(p_X',p_Y',p_Z')$
\begin{eqnarray}
\label{eqn-Bstep1}
p_X' &=& (p_X^2+p_Y^2)/p_S \\
p_Y' &=& 2 p_X p_Y / p_S \\
p_Z' &=& 2(1-p_X-p_Y-p_Z)p_Z/p_S \\
\label{eqn-Bstep2}
p_S &=& 1-2(p_X+p_Y)(1-p_X-p_Y),
\end{eqnarray}
where $p_S$ is the probability that a source EPR pair survives the step.
Note that half of the EPR pairs are target pairs and are always discarded after a B step.
\end{definition}

\begin{figure}
\centering
\includegraphics{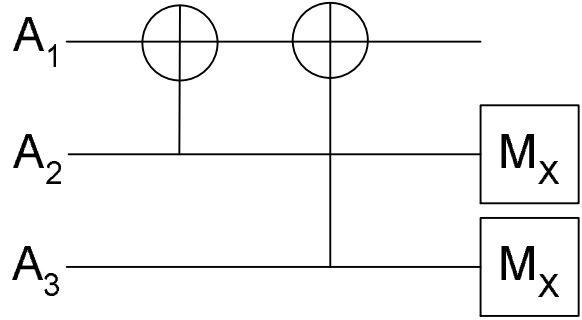}
\caption{\label{fig-pstep}
P step: Alice adds (module 2) her halves of three EPR pairs (the circuit of which is depicted) and Bob performs the same on his halves of the EPR pairs.
They keep the target EPR pair and discard the other two.
The $X$ measurements of the two source pairs do not need to be performed in a prepare-and-measure QKD.
}
\end{figure}

\begin{definition}[P step~\cite{Gottesman2003}]
A P step, shown in Fig.~\ref{fig-pstep}, operates on three EPR pairs randomly chosen, one target and two source pairs.
Alice and Bob perform a bilateral XOR on the target and a source pairs and then perform a second bilateral XOR on the target and the second source pairs.
The phase error syndrome is the $X$ measurements of the two source pairs, which is not needed in a prepare-and-measure QKD.
The target pair is kept for the next step.
The P step requires no communications between Alice and Bob and is really a classical circuit.
So, the P step commutes with the final $Z$ measurements of a QKD.
If the $Z$ measurements is performed before the P step, the P step is equivalent to XOR'ing three bits to generate one bit.
Suppose that initially the
EPR pairs are in the state $(p_X,p_Y,p_Z)$, a P step leads to a new state
\begin{eqnarray}
\label{eqn-Pstep0}
p_X' &=& 3 p_I^2 (p_X+p_Y) + 6 p_I p_X p_Z + 3 p_X^2 p_Y + p_X^3 \\
p_Y' &=& 6 p_I p_Y p_Z + 3 p_X (p_Y^2 + p_Z^2) + 3 p_Y p_Z^2 + p_Y^3 \\
p_Z' &=& 3 p_I (p_Y^2+p_Z^2) + 6 p_X p_Y p_Z + 3 p_Y^2 p_Z + p_Z^3 \\
\label{eqn-Pstep1}
p_I &=& 1-p_X-p_Y-p_Z,
\end{eqnarray}
where $p_I$ is the initial probability of no error.
Note that only one-third of the EPR pairs remain after a P step.
\end{definition}


The reduction from a EDP with B and P steps to BB84 is possible because these steps satisfy the ``no-branching (in $X$ operators) requirement'' in Gottesman-Lo's paper~\cite{Gottesman2003}.
Specifically, the decision of which EPR pairs to discard and which to retain only depends on the outcomes of $Z$ measurements, but not on the outcomes of $X$ measurements.
For BB84, Gottesman and Lo~\cite{Gottesman2003} showed that a sequence of five B steps, followed by six P steps, can give rise to a tolerable bit error rate of $18.9\%$.
Since $p_Y$ cannot be estimated in BB84, it is necessary to consider the worst-case value of $p_Y$ when determining the tolerable bit error rate.
They showed that $p_Y=0$ is the worst case for any sequence starting with a B step.
In this paper, we consider finding the highest tolerable bit error rate for SARG04 using Gottesman and Lo's technique.
We also prove the worst-case value of $p_Y$ for SARG04.





\section{Assumptions on the devices\label{sec-assumptions}}

In this section, we describe some assumptions we make in this paper.

First, note that Bob sometimes has a double click where he cannot determine the measurement outcome. This happens
because of the dark counts or detecting multi-photon. In this case, we impose Bob to take one of the bit values randomly
\cite{Inamori2005,Gottesman2004}. Thus, we can regard his measurement outcome as always stemming from the measurement
on a qubit state. This operation is so-called ``squash operation'' in~\cite{Gottesman2004}, which is a operation
mapping from a multi-photon state to a qubit state. Furthermore, we assume the measurement such that it can be
represented by the squash operation followed by a proper operations in a protocol. For instance, Bob's measurement
can be described by the squash operation followed by the rotations, the filtering operation and $Z$ basis measurement
in SARG04 protocol. We assume this model based on the squash operation in the whole paper.


In Section~\ref{sec-SARG04-deocy}, we will consider five types of imperfections in realistic QKD set-ups:
(i) the source is a laser source that generates a Poisson distribution of photon number state, (ii) there is loss in the optical fiber, (iii) Bob's detector is not completely efficient in declaring a detection event, (iv) Bob's detector may generate a false detection when there is no input, and (v) there is misalignment in Bob's detector.

Assuming the phase randomization, the single-mode laser source emits a pulse that is a classical mixtures
of the photon number states with a Poisson distribution:
\begin{eqnarray}
\sum_{i=0}^{\infty} \frac{\mu}{i!}e^{-\mu} \ket{i}\bra{i},
\end{eqnarray}
where $\mu$ is the mean photon number.

We quantify the loss in the optical fiber by the probability that an input photon is lost at the end of the transmission.
Let $\alpha$ in dB/km be the loss coefficient of the optical fiber and $l$ be the fiber length in km.
The probability that the input photon is not lost is equal to $10^{-\frac{\alpha l}{10}}$.

It is the case that Bob's detector fails to indicate the presence of an input photon.
The effect is similar to the transmission loss.
The probability that Bob's detector detects the presence of an input photon is defined as Bob's detection efficiency $\eta_{Bob}$.


Combining the loss in the quantum channel and the inefficiency of Bob's detector,
we have the overall transmission efficiency, $\eta$.
It is the probability that a photon is detected given that one has been sent, which is given by
\begin{eqnarray}
\eta &=& 10^{-\frac{\alpha l}{10}} \cdot \eta_{Bob}
\end{eqnarray}
When the input signal contains more than one photons, the signal is detected if at least one photon is detected.
Thus, the transmission efficiency for an n-photon signal is
\begin{eqnarray}
\eta_n &=& 1-(1-\eta)^n.
\end{eqnarray}

When there is no input to Bob's detector, there is a possibility that it generates a detection event.
This is due to the intrinsic detector's dark counts, the background spray, and the leakage from timing signals.
We denote the probability of this false detection event as $p_{dark}$.

We model the misalignment of the detectors by a rotation in the bases of Bob's projection measurements.
We will calculate the probabilities of inconclusive, correct, and incorrect results specifically for SARG04 using this model in Section~\ref{sec-SARG04-deocy}.

\section{SARG04 with one- and two-photon sources\label{sec:one-two-photon-source}}

In this section, we derive the lower and upper bounds of the tolerable bit error rates for SARG04 with two-way 
classical communications, where we consider using perfect one- and two-photon sources.

\subsection{Lower bounds with two-way communications}

\begin{figure}
\centering
\includegraphics[width=\figlength]{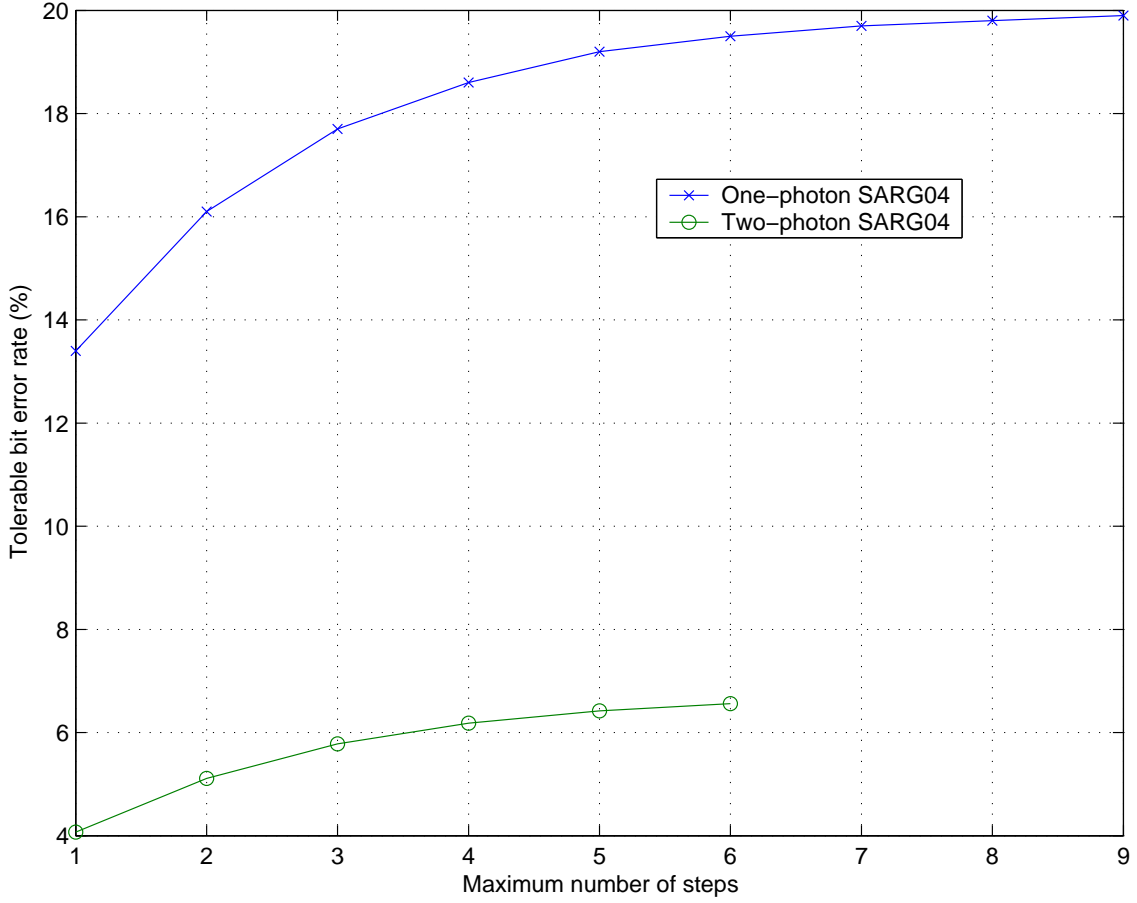}
\caption{\label{fig-single-1}The highest tolerable bit error rates with various maximum number of B or P steps allowed.
It turns out only B steps are used in all cases to achieve the highest bit error rates.
Even when a small number of B steps is used, the tolerable bit error rate increase quite substantially compared to the case where no B step is used.
Note that only up to 9 (6) steps are plotted for the single-photon (two-photon) case since sequences with more steps are less optimal.
}
\end{figure}

To determine the highest tolerable bit error rate, we would like to search for the sequence of B steps and P steps (introduced in Section~\ref{sec-prelim-two-way}) that, when followed by the one-way EDP with random CSS to correct bit and phase errors, gives a positive key generate rate for the bit error rate in question.
The sequence of B and P steps renders the initial state $(p_I,p_X,p_Y,p_Z)$ to another state $(p_I',p_X',p_Y',p_Z')$, which is then passed to the one-way protocol for producing almost perfect EPR pairs.
The key generation rate, based on the CSS protocol, is
\begin{eqnarray}
\label{eqn-CSSkeyrate}
R &=& 1- H(p_X'+p_Y')-H(p_Z'+p_Y') .
\end{eqnarray}
Note that we have ignored the mutual information between the bit and phase errors for simplicity of analyses.
For the single-photon case of SARG04, the initial state is $p_X=e_b -a$, $p_Z=\frac{3}{2}e_b-a$, $p_Y=a$, where Alice and Bob can estimate $e_b$ but not $a$.
Thus, for the purpose of determining the highest tolerable bit error rate, we consider the worst-case value of $a$ for a given $e_b$ and a given sequence such that the initial state with this value will lead to the smallest key generation rate.
A proof of this for BB84 was given in~\cite{Gottesman2003}.
Here we adapt their proof to SARG04 and have the following theorem:
\begin{theorem}
\label{thm-worst-SARG04-single}
For an initial state of $p_X=e_b -a$, $p_Z=\xi e_b-a$, $p_Y=a$, where $\xi \in \field{R} \geq 1$ is some constant,
the key generation rate as given in \myeqnref{eqn-CSSkeyrate} is an increasing function of $a$ for a fixed $e_b$ and a fixed sequence of B steps and P steps starting with a B step, under the following conditions:

(i) $e_b < \frac{1+4 a}{2 (1+\xi)} \forall a $ in the valid range, and

(ii) $e_b< \frac{1}{2\xi}$.
\end{theorem}
\begin{proof}
See Appendix~\ref{app-worst-SARG04-single}.
\end{proof}
Note that Theorem~\ref{thm-worst-SARG04-single} is a simple generalization of the result in Appendix~III of~\cite{Gottesman2003}.
For the single-photon case, we apply Theorem~\ref{thm-worst-SARG04-single} with $\xi = \frac{3}{2}$.
Given the valid range of $a$ being $[e_b/2,e_b]$, we have the following:
\begin{corollary}
The worst-case for single-photon SARG04 is $a=\frac{e_b}{2}$.
\end{corollary}
We have written a simple computer program in Mathematica to calculate the evolution of the diagonal elements of the marginal density matrix of the EPR pairs shared by Alice and Bob under sequences of B and P steps using \myeqnrefs{eqn-Bstep1}-(\ref{eqn-Pstep1}).
With $a=\frac{e_b}{2}$, we exhaustively searched for the step sequence with 15 B/P steps or less that can tolerate the highest bit error rate.
For each sequence, we searched for the highest initial value of $e_b$ that gives rise to a positive key generation rate given by \myeqnref{eqn-CSSkeyrate}.
We conclude that $e_b=19.9\%$ is tolerable with nine B steps.
We can easily check that this value of $e_b$ satisfies the two conditions of Theorem~\ref{thm-worst-SARG04-single}.
Since in each B step, Alice and Bob discard at least half of the EPR pairs that have survived so far, a protocol with nine B steps leaves only a small number of EPR pairs at the end of the protocol.
Thus, a sequence with nine B steps may not be efficient in practice.
Therefore, we consider the highest tolerable bit error rates with various maximum numbers of steps allowed, as shown in Fig.~\ref{fig-single-1}.
As can be seen, even a protocol with two B steps is able to tolerate a bit error rate of 16.1\%, which is a great improvement from that of one-way protocols ($10.95\%$ from~\cite{Tamaki2005,Branciard2005}).

%
%
%
%
\begin{matlabcode}{
plot([1:9],[13.4,16.1,17.7,18.6,19.2,19.5,19.7,19.8,19.9],'x-',[1:6],[4.07,5.11,5.78,6.18,6.42,6.56],'o-')
grid on
legend('One-photon SARG04','Two-photon SARG04')
xlabel('Maximum number of steps')
ylabel('Tolerable bit error rate (
}\end{matlabcode}

We now consider two-photon SARG04, whose density matrix satisfies \myeqnref{eqn-two-1}.
Since $p_Z \leq x e_b + g(x) -a , \forall x$, we can minimize the right-hand side over $x$ to find the worst-case $p_Z$.
Substituting in the minimizing $x$ gives us the initial state
\begin{eqnarray}
p_X&=&e_b-a \nonumber \\
\label{eqn-two-2}
p_Z &=& \frac{1}{2}- \frac{1}{2\sqrt{2}} (1-3e_b) + \sqrt{\frac{1-(1-3e_b)^2}{24}} -a  \\
p_Y&=&a, \nonumber
\end{eqnarray}
where $0 \leq a \leq e_b$.
Since $p_Z$ can be written as $p_Z=\xi e_b - a$ for some $\xi \geq 1$ and for a fixed $e_b$,
we can invoke Theorem~\ref{thm-worst-SARG04-single} to arrive at the following:
\begin{corollary}
The worst-case for two-photon SARG04 is $a=0$.
\end{corollary}
In the worst case, we found that $e_b=6.56\%$ is tolerable with six B steps for two-photon SARG04.
This is greater than the tolerable bit error rate of $2.71\%$ using one-way communications~\cite{Tamaki2005}.
The highest tolerable bit error rates with various maximum numbers of steps allowed for the two-photon is also shown in Fig.~\ref{fig-single-1}.
As can be seen, even when a smaller number of B steps is used,
the tolerable bit error rate increase quite substantially compared to the case where no B step is used.

The steps for the SARG04 protocol with a $\nu$-photon source ($\nu=1,2$) involving B steps are similar to the one-way SARG04 protocol in Section~\ref{sec-SARG04} and are as follows:
\begin{enumerate}
\item[1-6.] Same as that in one-way SARG04.
\item[7.] B step: Alice randomly divides the bits into pairs and informs this to Bob.
They separately compute the parity for each pair and compare their results with each other.
If they have the same parity for a pair, they keep one bit and discard the other bit of the pair;
otherwise, both bits are discarded.
This step is repeated as many times as needed.

\item[8.] They perform bit error correction and privacy amplification on the remaining bit string using the revised bit error rate.
\end{enumerate}

\subsection{Upper bounds with two-way communications}

An upper bound for single-photon SARG04 with one-way communications was provided in~\cite{Branciard2005}.
This upper bound of $14.9\%$ is lower than our lower bound of $19.9\%$ with two-way communications.
In other words, as far as the single-photon component is concerned, SARG04 with two-way classical communications can tolerate a higher bit error rate than SARG04 with only one-way classical communications.
A similar behaviour was previously found in BB84~\cite{Gottesman2003}.
Here, we will investigate the upper bound with two-way communications for both single-photon and two-photon in SARG04.

To arrive at an upper bound, we note that security cannot be established between Alice and Bob if there is no entanglement shared between them~\cite{Curty2004}.
Specifically, when the density matrix of Alice and Bob is separable, i.e. $\rho_{AB} = \sum_i \rho_{A,i} \otimes \rho_{B,i}$, then there is no entanglement.
One result from BDSW~\cite{BDSW1996} is that, for a bipartite state with a density matrix of the form
\begin{eqnarray}
\rho &=& p_1 \ket{\Phi^+}\bra{\Phi^+} + p_2 \ket{\Phi^-}\bra{\Phi^-}
+ p_3 \ket{\Psi^+}\bra{\Psi^+} + p_4 \ket{\Psi^-}\bra{\Psi^-} ,
\end{eqnarray}
if none of the probabilities $p_1,\dots, p_4$ is greater than $\frac{1}{2}$, then $\rho$ can be written as a mixture of separable states and thus no entanglement exists.
Using this idea, we may find the bit error rate with which the Bell diagonal elements of our density matrices of SARG04 in \myeqnref{eqn-single-densitymatrix} and \myeqnref{eqn-two-densitymatrix} are all no greater than $\frac{1}{2}$.
We may imagine $e_b$ to be small initially, in which case $p_I$ is close to unity and $p_X$, $p_Y$, and, $p_Z$ are close to zero.
Then, we gradually increase $e_b$ until $p_I$ goes down to $\frac{1}{2}$.
Although the BDSW idea applies only to Bell-diagonal density matrix and our density matrices may not be Bell diagonal, we can still apply the BDSW idea to our case since whether the off-diagonal terms are zero or not has no bearing on the B steps, the P steps, the CSS error correction, and the CSS privacy amplification in our protocol.
In other words, our entanglement distillation method does not extract entanglement from the off-diagonal terms.
Thus, we may safely regard our density matrices as Bell-diagonal.

For single-photon SARG04, setting $p_I=\frac{1}{2}$ gives $e_b=\frac{1}{5}+\frac{2a}{5}$.
Given the valid range of $a$, this suggests that $e_b$ is between $\frac{1}{4}$ and $\frac{1}{3}$.
Eve would like to cause the error rate as low as possible.
But she may not be able to choose $a$ freely to induce an error rate of $1/4$, since $a$ is an parameter influenced by her and is not in her complete control in any attack strategy by her.
Thus, without any reference to a specific attack strategy, the value of $a$ (and the upper bound on $e_b$) cannot be specified.
Therefore, we focus on specific intercept-and-resend strategies to determine specific values of $a$ and an upper bound on $e_b$.

In an intercept-and-resend attack, Eve captures and measures the photon sent by Alice to Bob.
She then sends another photon with the polarization depending on the measurement result to Bob.
Certainly, no entanglement exists between Alice and Bob, since Bob's photon was created by Eve.
In a simple intercept-and-resend attack, Eve performs a photon polarization measurement with a basis randomly chosen from two bases.
The first basis consists of $\ket{\varphi_0}$ and $\ket{\varphi_2}$, while the second consists of $\ket{\varphi_1}$ and $\ket{\varphi_3}$.
After the measurement, Eve sends the resultant state to Bob.
This particular attack causes an error rate of $\frac{1}{3}$.
The fact that this is at the high end of the range $[\frac{1}{4},\frac{1}{3}]$ prompts us to search for a more sophisticated intercept-and-resend attack.

\begin{definition}[General POVM attack]
A general POVM attack is an individual intercept-and-resend attack by Eve who captures every transmission from Alice (each which may consist of one or more photons), performs an arbitrary POVM measurement on each transmission independently, and sends an arbitrary state to Bob depending on the measurement outcome.
The POVM is arbitrary and can be represented by $J+1$ elements, $\{M_{\text{vac}},M_i, i=[0,\ldots,J-1]\}$, with $M_{vac}+\sum_i M_i=I$.
For the outcome corresponding to $M_{\text{vac}}$, Eve sends vacuum to Bob,
whereas, for outcome $i$, she sends an arbitrary state $\ket{\sigma_i}$ to Bob.
\end{definition}
We consider Eve launching such a general POVM attack for the SARG04 one-photon case and two-photon case.
We want to optimize over $M_{vac}$, $M_i$, and $\ket{\sigma_i}$ so that Eve induces the lowest possible bit error rate, hoping to achieve a rate smaller than $1/3$ caused by the simple attack described above for the one-photon case.
Unfortunately, for the one-photon case, even with such a great freedom to choose the POVM and the states sent, this attack cannot do better than the simple attack.
\begin{theorem}
\label{thm-SARG04-generalattack1}
For single-photon SARG04, the smallest bit error rate $e_b$ caused by Eve using a general POVM attack is $\frac{1}{3}$.
\end{theorem}
\begin{proof}
See Appendix~\ref{app-SARG04-generalattack}.
\end{proof}

On the other hand, for two-photon SARG04, it is not trivial to consider intercept-and-resend attack and thus
we only consider a general POVM attack.
\begin{theorem}
\label{thm-SARG04-generalattack2}
For two-photon SARG04, the smallest bit error rate $e_b$ caused by Eve using a general POVM attack is $\frac{3-\sqrt{2}}{7} \approx 22.65\%$.
Moreover, a POVM that gives rise to this minimum bit error rate is
\begin{eqnarray}
M_m &=& P(\lambda_+ \ket{\varphi_m} \ket{\varphi_m} + \lambda_- \ket{\varphi_{m+2}} \ket{\varphi_{m+2}})
\: , m=0,\ldots,3 \\
M_{\text{vac}} &=&
P(\ket{\varphi_0} \ket{\varphi_2} - \ket{\varphi_2} \ket{\varphi_0})/2 \\
&=&P(\ket{\varphi_3} \ket{\varphi_1} - \ket{\varphi_1} \ket{\varphi_3})/2
\end{eqnarray}
where $\lambda_{\pm}=(\pm2+\sqrt{2})/4$, $P(\ket{\Phi})=\ket{\Phi}\bra{\Phi}$ is a projection operator associated with a pure state $\ket{\Phi}$, and the subscript in $\varphi_{m+2}$ is taken in {\em modulo} $4$.
Eve sends $\ket{\varphi_m}$ to Bob when the measurement outcome is $m \in [0,3]$.
Note that $M_{\text{vac}}$ never occurs, since the four states sent by Alice, $\ket{\varphi_m} \ket{\varphi_m}, m \in [0,3]$, are orthogonal to the state $M_{\text{vac}}$ projects onto.
\end{theorem}
\begin{proof}
See Appendix~\ref{app-SARG04-generalattack}.
\end{proof}

\subsection{Comparison with BB84 in depolarizing channels}

We compare the lower and upper bounds with two-way communications of SARG04 and of BB84 by assuming that the eavesdropping is realized by a depolarizing channel.
A depolarizing channel evolves an $\nu$-photon input $\rho^{\otimes \nu}$ to $(1-\frac{4p}{3}) \rho^{\otimes \nu} + \frac{4p}{3} ( I/2)^{\otimes \nu}$ with a depolarizing rate $p$.
For SARG04, the depolarizing rate $p$ is related to the bit error rate $e_b$ by $e_b = 4p/(3+4p)$, whereas, for BB84, $e_b=2p/3$.
Using these formulas, we see that SARG04 is secure up to $p \approx 18.6\%$ for one-photon and $p \approx 5.27\%$ for two-photon, and BB84 is secure up to $p \approx 28.35\%$~\cite{Gottesman2003} with two-way communications.
For the upper bounds, SARG04 is insecure beyond $p=3/8$ for one-photon and $p=3(2-\sqrt{2})/8 \approx 22.0\%$ for two-photon, and BB84 is insecure beyond $p=3/8$~\cite{Gottesman2003}.





\section{\label{sec-SARG04-deocy}SARG04 with realistic sources using decoy}


With a realistic phase-randomized laser source, the output pulses are classical mixtures of the photon number states with a Poisson distribution.
In this section, we consider using the decoy method of~\cite{Lo2005} to operate SARG04 securely with a realistic source.
With this particular decoy method,
the mean photon number of the laser source when emitting the decoy states varies over infinitely many values, in order to estimate the statistics for the decoy states.
Works in~\cite{Lo2004,Ma2005b,Wang2005a,Wang2005b,Harrington2005} analyzed practical decoy schemes with only a few decoy states.
Here, we consider applying the infinite-decoy idea to SARG04 for the simplicity of analyses.

The steps for the SARG04 protocol with decoy states are as follows:
\begin{enumerate}
\item Alice randomly chooses the locations of the decoy states and the signal states.
\item For the decoy states, Alice adjusts the power of the laser to have a random mean-photon number $\mu$ and she records this value of $\mu$.
For signal states, Alice operates the laser at a fixed mean-photon number.
\item Alice randomly chooses one of the four sets and sends one of the two states in the set to Bob.
\item Bob performs the polarization measurement using one of the two bases randomly. If his detector fails to click, then he broadcasts this fact, and Alice and Bob discard all the corresponding data.
\item Alice announces the sets of states for both decoy and signal states to Bob.
She also announces the locations of the decoy states, their values of $\mu$, and their states.
\item Bob, based on the information on the sets of states, broadcasts which bits are conclusive or not.
\item For all the decoy states having the same $\mu$, Bob estimates $Q_\mu$ by taking the ratio of the number of conclusive events to the total number of conclusive, inconclusive, and no-detection events.
He estimates $E_\mu$ by taking the ratio of the number of incorrect conclusive events to the total number of conclusive events.
\item Bob then estimates $e_1$ and $e_2$ based on $Q_\mu$'s and $E_\mu$'s over all values of $\mu$'s.
\item If both of $e_1$ and $e_2$ are too high, they abort the protocol.
\item Alice and Bob discard all events concerned with inconclusive and all decoy states.
\item They perform bit error correction on the remaining bit string and apply privacy amplification.
\end{enumerate}

In this section, we analyze the key generation rate of this protocol under
the same situation as was considered in~\cite{Lo2005}, in which
(i) the source is a phase randomized coherent source, (ii) there is loss in the optical fiber, (iii) Bob's detection is not completely efficient in declaring a detection event, (iv) there are dark counts, and (v) there is misalignment in Bob's detector.
We first develop a specific detector error model for SARG04, which is then be used to formulate the yield and the error rate equations for SARG04.
With the yields and the error rates, we can compute the achievable key-generation rates.

\subsection{Model for detector errors in SARG04}

We consider a specific error model for detections in SARG04.
We have chosen this model because it is also a simple model for explaining errors in BB84 and thus would provide a reasonable performance comparison with BB84.
In the decoy paper for BB84~\cite{Lo2005}, they used the Gobby-Yuan-Shields (GYS)~\cite{Gobby2004} experimental results to characterize the probability of detector error in BB84, denoted by $e_{detector}$.
The value of this probability is specific to the setup in the GYS experiment which is for BB84.
Although an experimental setup for SARG04 might be the same as that for BB84 (since their quantum phases are the same),
their interpretations of errors are different
and thus there is no reason to believe that the error probabilities describing both setups are exactly the same.
Nevertheless, in order to facilitate a reasonable comparison between SARG04 and BB84, we attribute the probability of detector error to a rotation of the detector by a small angle.
Specifically, we model the misalignment of the detectors by a rotation of angle $\theta$ in the two projection measurements at Bob's side.
Using the same model for both SARG04 and BB84, we can compare their results on a common ground.
For SARG04, we can calculate the probabilities of getting the inconclusive, incorrect, and correct outcomes for each of the four bases.
\begin{figure}
\centering
\includegraphics[width=\figlength]{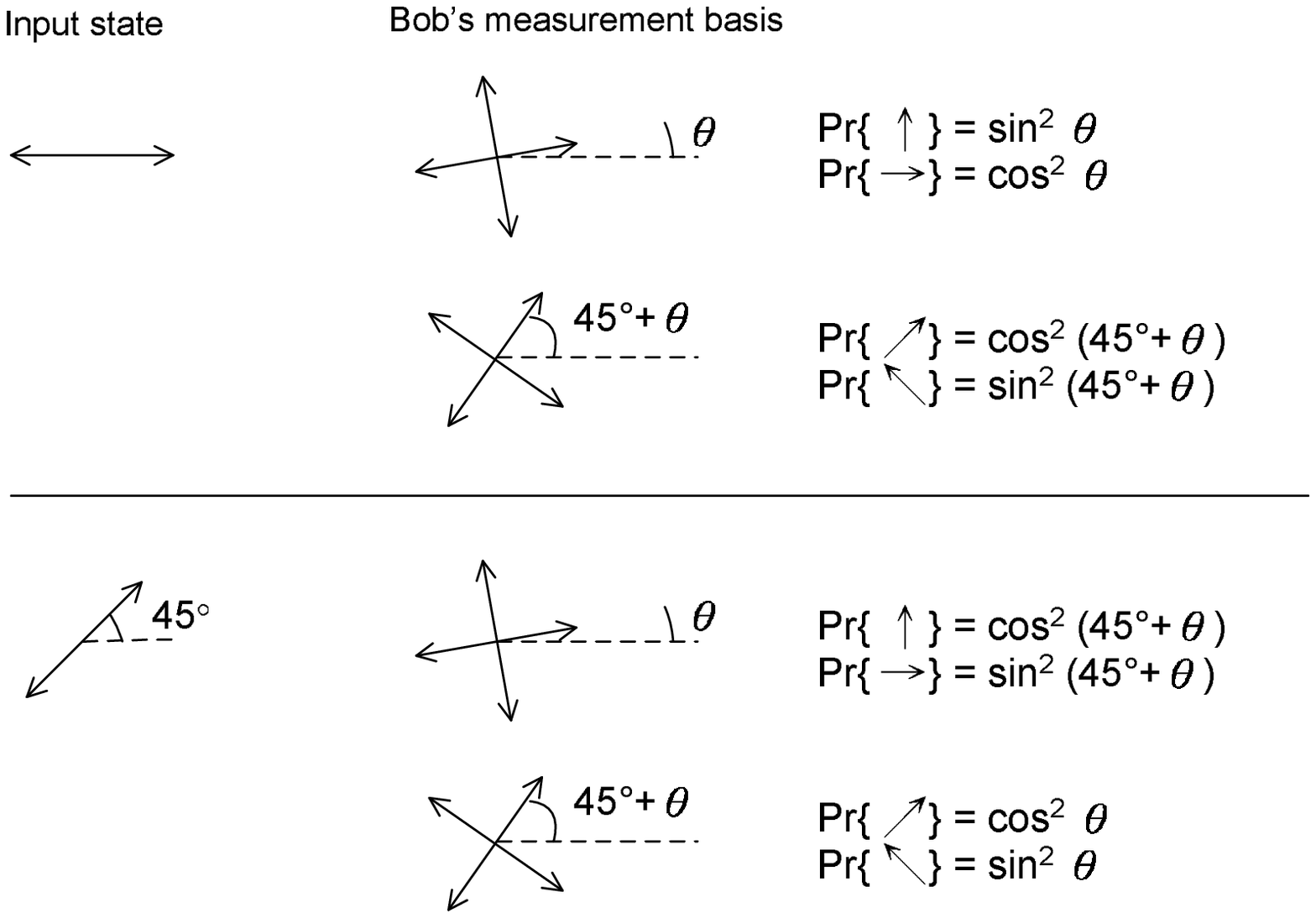}
\caption{\label{fig-decoy-1}
Calculation of measurement probabilities with misalignment of the detectors for the basis $\{\leftrightarrow,\nearrow\}$.
The misalignment is modeled by a rotation of the two measurement bases, $\theta$.
For this basis, the conclusive results are $\updownarrow$ and $\nwarrow$.
}
\end{figure}
For example, Fig.~\ref{fig-decoy-1} shows the calculation for the basis $\{\leftrightarrow,\nearrow\}$.
In the end, we conclude that given a successful detection event at Bob's detector,
$Pr\{conclusive\}=\frac{\sin^2(\theta)}{2}+\frac{1}{4}$,
$Pr\{incorrect\}=\frac{\sin^2(\theta)}{2}$, and
$Pr\{correct\}=\frac{1}{4}$.
For BB84, the probability of detection error can easily be seen to be $\sin^2(\theta)$.
Similarly, we perform the same calculations when Bob detects a vacuum state and a dark count occurs.
We arrive at
$Pr\{inconclusive\}=1/2$, $Pr\{correct\}=1/4$, and $Pr\{incorrect\}=1/4$.
These probabilities are used later in the calculations of the yields and the bit error rates for SARG04.


\subsection{Key generation rate using decoy}

Recall that the key generation rate for SARG04 with decoy is
\begin{eqnarray}
\label{eqn-decoyrate1}
R_{\text{\tiny SARG04}} \geq -Q_\mu f(E_\mu) H_2 (E_\mu) + Q_1 [1-H_2(Z_1|X_1)] + Q_2 [1-H_2(e_{p,2})],
\end{eqnarray}
where
the subscript $\mu$ denotes the mean photon number for the signal states,
$Q_\mu$ is the gain
of the signal states, $E_\mu$ is the QBER of the signal states,
$Q_j$ and $e_{p,j}, (j=1,2)$ are the gains and the phase error rates of the single-photon states $(j=1)$ and the two-photon states $(j=2)$,
$Z_1$ and $X_1$ are random variables characterizing the phase and bit errors for the single-photon states (see Section~\ref{sec:Preliminaries} for definition),
$f(x)$ is the error correction efficiency as a function of error rate, and
$H_2(x)$ is the binary entropy function.

We note that both single-photon states and two-photon states have positive contributions to the key generation rate, in contrast to BB84, the key generation rate of which has only the single-photon-state contribution.
Also, since there is mutual information between the bit and phase errors for the single-photon case, we have included this contribution to the key generation in \myeqnref{eqn-decoyrate1}.
The parameters $Q_\mu$ and $E_\mu$ in \myeqnref{eqn-decoyrate1} can be estimated through public communications.
The phase error rates $e_{p,1}$ and $e_{p,2}$ can be estimated respectively from the bit error rates $e_1$ and $e_2$ (using the relations in \myeqnref{eqn-single-densitymatrix} and \myeqnref{eqn-two-densitymatrix} with the worst-case values of $a=e_1/2$ and $a=0$ respectively).
The bit error rates $e_1$ and $e_2$, along with $Q_1$ and $Q_2$, can in turn be estimated using the decoy state idea.
In what follows, we derive the formulas for these parameters for SARG04, and thus, using these parameters, we can determine the key generation rate using \myeqnref{eqn-decoyrate1}.

\subsection{Yields and bit error rates}
%
We now determine the yields and the bit error rates of the transmitted qubits for SARG04.
Using the definition of the yield in \myeqnref{eqn-def-yield},
the yield for SARG04 is
\begin{eqnarray}
\label{eqn-SARG04Yn}
Y_{n,\text{\tiny SARG04}} &=&
\eta_n \left( \frac{e_{detector}}{2}+\frac{1}{4} \right) + (1-\eta_n) p_{dark} \frac{1}{2},
\end{eqnarray}
where $e_{detector}=\sin^2(\theta)$.
The fraction of $\frac{1}{4}$ corresponds to the probability of getting a conclusive result.
Compared to the yield for BB84 in \myeqnref{eqn-BB84Yn}, we see that the yield stemmed from the signal for SARG04 is approximately half of that for BB84.
On the other hand, the yields stemmed from the dark count are the same for SARG04 and BB84.
Similarly, for the bit error rate,
\begin{eqnarray}
\label{eqn-SARG04en}
e_{n,\text{\tiny SARG04}} &=&
\left[ \eta_n \frac{e_{detector}}{2} + (1-\eta_n) p_{dark} \frac{1}{4} \right] / Y_{n,\text{\tiny SARG04}} .
\end{eqnarray}
%
%
Thus, the overall gain and the overall QBER for the coherent state $\ket{\sqrt{\mu}}$ are, respectively,
\begin{eqnarray}
Q_{\mu,\text{\tiny SARG04}} 
&=& \frac{1}{2} p_{dark}e^{-\eta \mu} + \left( \frac{e_{detector}}{2} + \frac{1}{4} \right) ( 1- e^{-\eta \mu}) , \text{ and}
\end{eqnarray}
\begin{eqnarray}
E_{\mu,\text{\tiny SARG04}} 
&=& \left[ \frac{1}{4} p_{dark}e^{-\eta \mu} + \frac{e_{detector}}{2} ( 1- e^{-\eta \mu}) \right] / Q_{\mu,\text{\tiny SARG04}}.
\end{eqnarray}
Using these formulas for the error rates and the gains, we can compute the key generation rate for SARG04 with one-way decoy using \myeqnref{eqn-decoyrate1}.

\subsection{Simulations}

\begin{figure}
\centering
\includegraphics[width=\figlength]{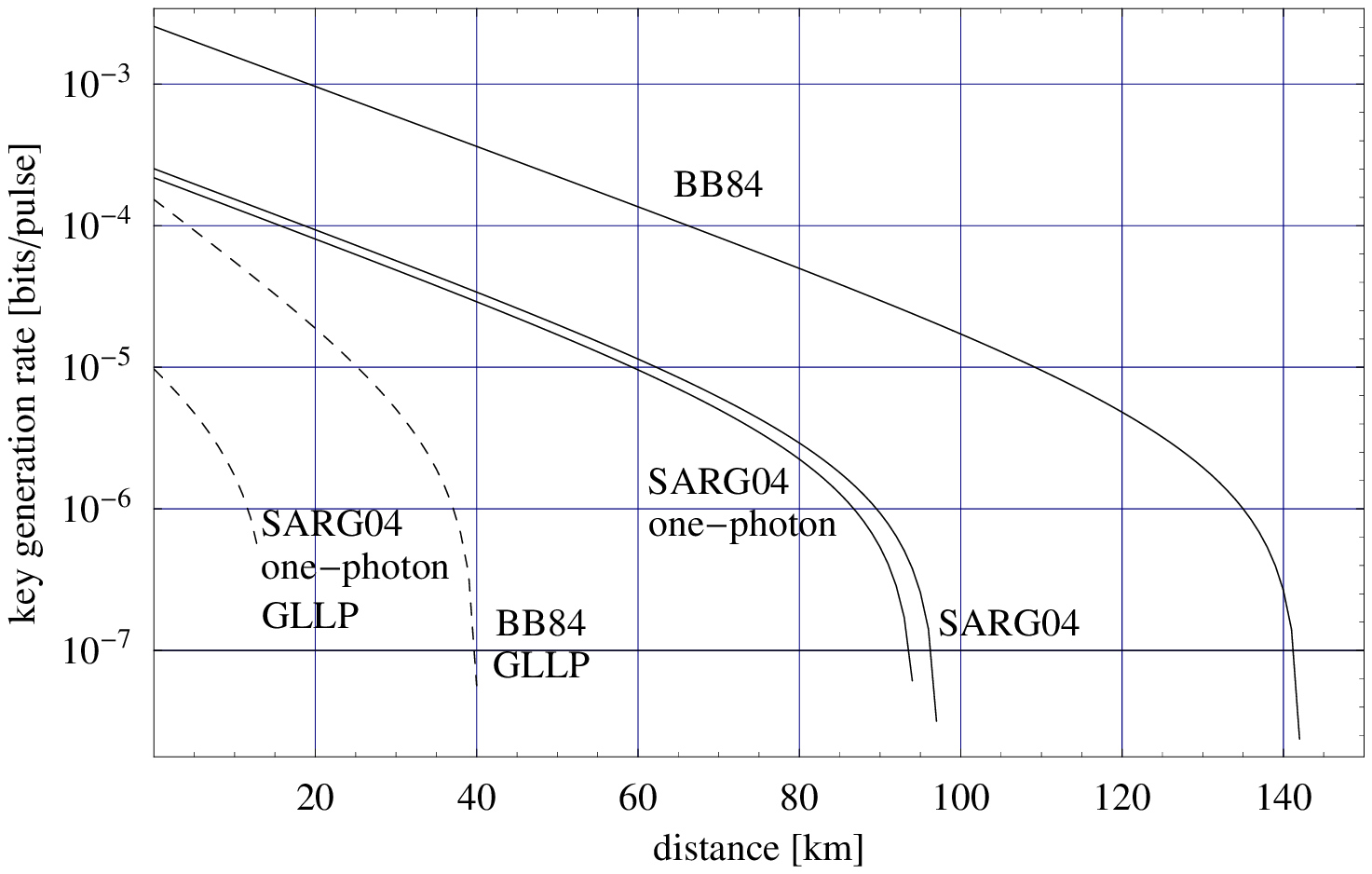}
\caption{\label{fig-decoygraph1}
Simulation using the GYS parameters listed in Table~\ref{table-sim-param} and $f(E_\mu)=1.22$,
for both SARG04 and BB84.
We compare the key generation rates of SARG04 and BB84 using decoy states (solid curves) and without using decoy states (dashed curves).
The optimal mean photon numbers, $\mu$, for all curves are used at all distances.
Two curves of SARG04 using decoy are plotted, one with both single- and two-photon contributions and the other with only single-photon contributions.
Also, curves of single-photon SARG04 and of BB84 using GLLP without decoy are plotted.
The maximal secure distance is $97.2$km for SARG04 and $141.8$km for BB84.
However, the upper bounds for SARG04 and for BB84 are exactly the same, namely, $207.68$ km.}
\end{figure}%
Fig.~\ref{fig-decoygraph1} compares the key generation rates of SARG04 and BB84, both using the one-way infinite-decoy method.
For this simulation, we take $f(E_\mu)=1.22$ for simplicity and use the parameters from the experiments by Gobby {\it et al.}~\cite{Gobby2004} as shown in Table~\ref{table-sim-param}.
We assumed that the detectors in both cases are rotated by the same angle in our model.
The optimal mean photon numbers, $\mu$, for SARG04 and BB84 are used at all distances.
Two curves of SARG04 using decoy are plotted, one with both single- and two-photon contributions and the other with only single-photon contributions.
Comparing these two curves, it can be seen that the two-photon part has a small contribution to the key generation rates at all distances.
Also, curves of single-photon SARG04 and of BB84 using GLLP without decoy are plotted.
We see that, by using decoy, higher key generation rates and longer secure distance can be achieved.
A similar behaviour for BB84 was shown in~\cite{Lo2005}.
We note that the key generation rate for BB84 with GLLP in Fig.~1 of \cite{Lo2005} is smaller than ours.
This is because we used the optimal $\mu$ for all distances in Fig.~\ref{fig-decoygraph1} while $\mu$ proportional to $\eta$ was used in \cite{Lo2005}.
The maximal secure distance for SARG04 using decoy is $97.2$km, compared to $141.8$km for BB84.
The upper bound of the distance in SARG04 can be determined by finding the distances corresponding to $e_1=\frac{1}{3}$ and to $e_2=0.2265$;
they are, respectively, $207.68$ km and $201.43$ km.
Thus, the upper bound of the distance is $207.68$ km, at which the two-photon part is not secure but the single-photon part is.
Interestingly, this bound of $207.68$ km is exactly the same as the upper bound for BB84~\cite{Lo2005}.
It can be shown analytically that setting $e_1=\frac{1}{3}$ for the SARG04 case and setting $e_1=\frac{1}{4}$ for the BB84 case both give the same formula for $\eta_1$, specifically, $\eta_1=\frac{p_{dark}}{1-4e_{detector}+p_{dark}}$.
(The formulas for $e_n$ and $Y_n$ of BB84 are of course different from that of SARG04.)
\begin{table}
\centering
\begin{tabular}{|c|c|c|c|c|}
\hline
Wavelength [nm] & $\alpha$ [dB/km] & $\eta_{Bob}$ & $e_{detector}$ & $p_{dark}$ \\ \hline
$1550$ & $0.21$ & $4.5\%$ & $3.3\%$ & $1.7 \times 10^{-6}$ \\ \hline
\end{tabular}
\caption{\label{table-sim-param}Simulation parameters from  Gobby-Yuan-Shields (GYS) experiments~\cite{Gobby2004}.}
\end{table}
\begin{figure}
\centering
\includegraphics[width=\figlength]{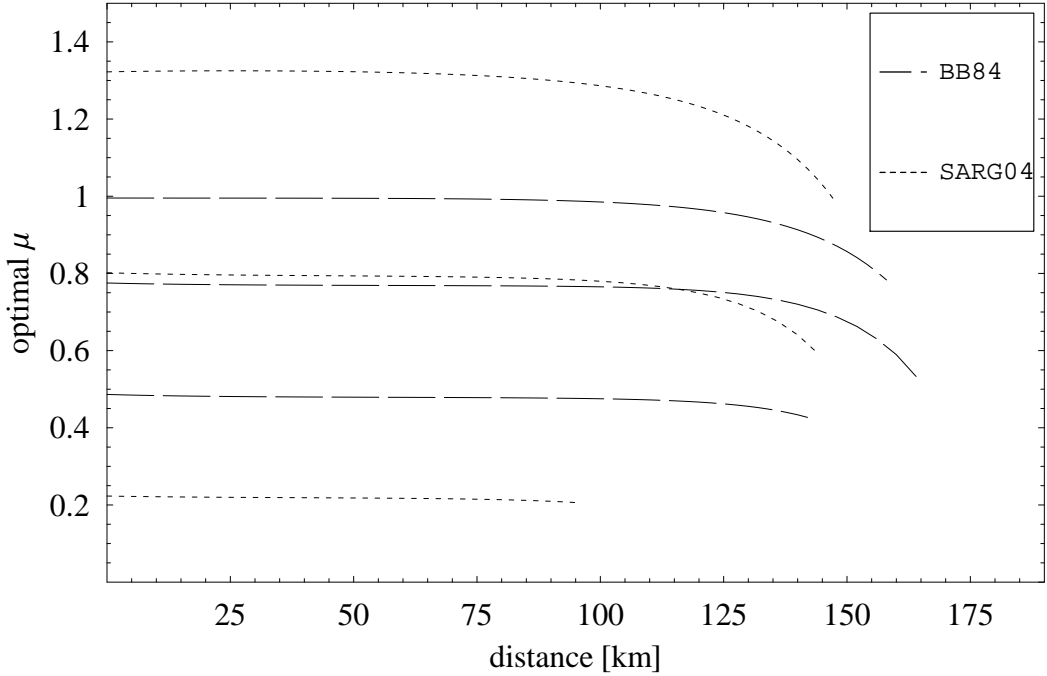}
\caption{\label{fig-decoy-mu}The optimal $\mu$'s for achieving the highest key generation rate at each distance for SARG04 using \myeqnref{eqn-decoySARG04rate1} and for BB84 using \myeqnref{eqn-decoyBB84rate1}.
Three sets of parameters are plotted.
The bottom two, middle two, and top two curves for BB84 and SARG04 used the same parameters except for $e_{detector}$ (which are $0.033$, $0.01$, and $0.0001$, respectively).
The other common parameters are listed in Table~\ref{table-sim-param} and $f(E_\mu)=1.22$.
When the misalignment of the detector is large (i.e. large $e_{detector}$) as in the bottom two curves, BB84 uses a laser with a higher optimal mean photon number than SARG04.
When the misalignment becomes smaller as in the top two curves, the situation is reversed; SARG04 operates optimally with a higher $\mu$ than BB84 does.
Also, note that the optimal $\mu$ for SARG04 can be higher than one.
}
\end{figure}
The optimal $\mu$'s for achieving the highest key generation rate at each distance for SARG04 and BB84 using decoy are plotted in Fig.~\ref{fig-decoy-mu}.
We can see that, when the misalignment of the detector is large (i.e. large $e_{detector}$), the optimal mean photon number for BB84 is higher than that of SARG04.
On the other hand, when the misalignment is small, the optimal $\mu$ of SARG04 is higher at short and medium distances.
In addition, the optimal $\mu$ of SARG04 can be higher than one in this case.
This is reasonable since at short or medium distances, the bit error rate is not high and thus the key contribution from the two-photon part in SARG04 is relatively high;
on the other hand, at long distances, the two-photon contribution is relatively small.
Since the optimal $\mu$ for SARG04 and BB84 is approximately constant for a large range of distances, the key generation rates for both of SARG04 and BB84 are in the order of $O(\eta)$.

\begin{figure}
\centering
\includegraphics[width=\figlength]{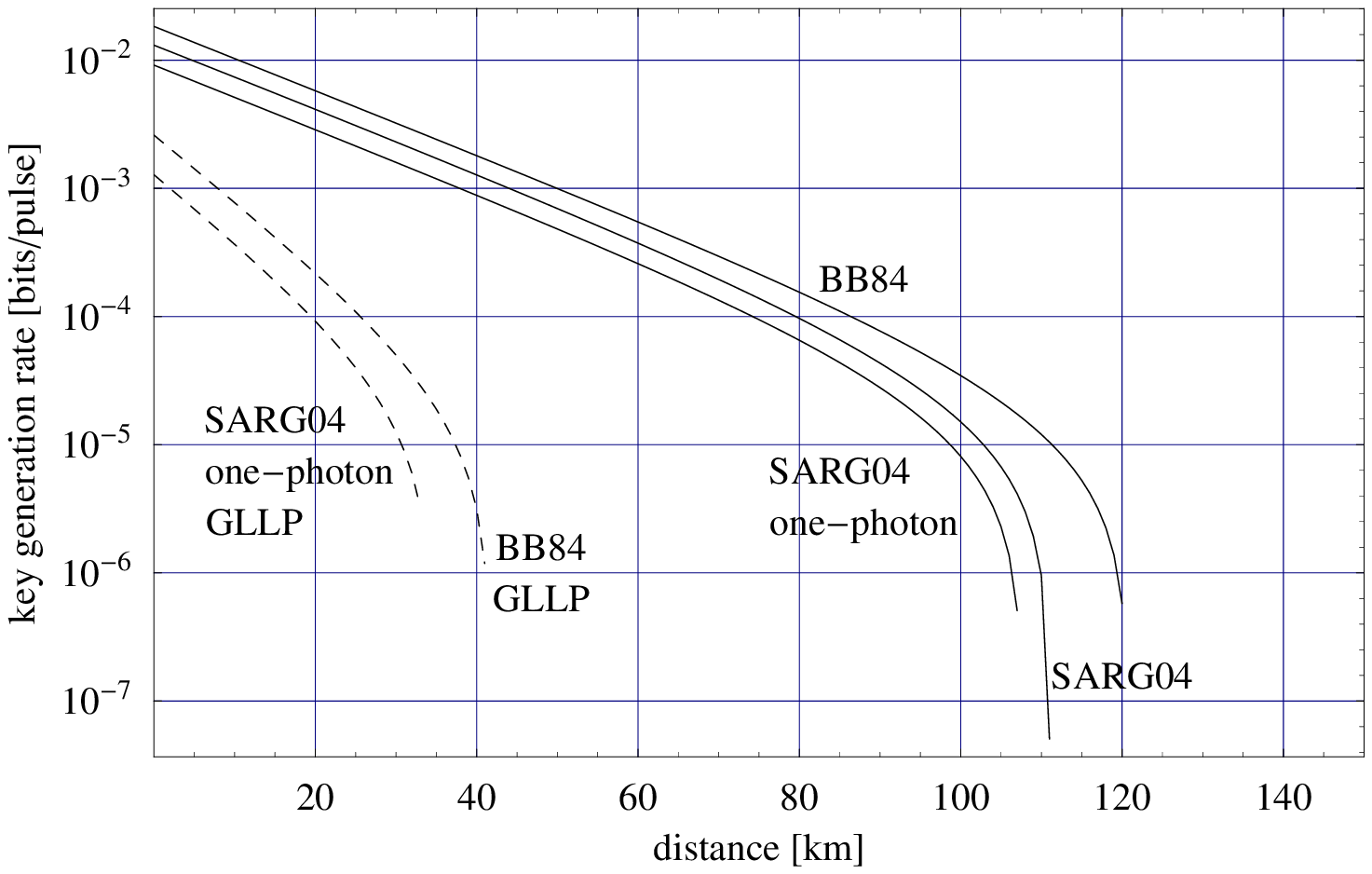}
\caption{\label{fig-decoygraph1b}
Simulation using
$\alpha=0.25$, $\eta_{Bob}=0.1$, $e_{detector}=0$, $p_{dark}=10^{-5}$ (from~\cite{Branciard2005}), and $f(E_\mu)=1$,
for both SARG04 and BB84.
We compare the key generation rates of SARG04 and BB84 using decoy states (solid curves) and without using decoy states (dashed curves).
The optimal mean photon numbers, $\mu$, all curves are used at all distance.
Two curves of SARG04 using decoy are plotted, one with both single- and two-photon contributions and the other with only single-photon contributions.
Also, curves of single-photon SARG04 and of BB84 using GLLP without decoy are plotted.
}
\end{figure}
Fig.~\ref{fig-decoygraph1b} shows the simulation using the parameters from Fig.~4 of~\cite{Branciard2005}.
Our result shows that, under our assumption that Eve may perform the most general attack, BB84 is able to achieve both a higher secret key rate and a greater secure distance than SARG04, whereas,
under the assumption considered by~\cite{Branciard2005} that Eve may only perform incoherent attacks, they observed the reverse phenomenon in Fig.~4 of their paper (i.e. SARG04 has a higher key rate and greater distance than BB84).
Another difference between our result and that of~\cite{Branciard2005} is that we also consider contributions from the two-photon part.

In both Figs.~\ref{fig-decoygraph1} and \ref{fig-decoygraph1b}, there are gaps between the one-photon SARG04 curves and the BB84 curves whether or not decoy is used.
These gaps are mainly due to the decrease in the gain $Q_n$ and the increase in the bit error rate $e_n$ in SARG04 relative to BB84.
We can see this as follows.
By comparing the yields of SARG04 in \myeqnref{eqn-SARG04Yn} and of BB84 in \myeqnref{eqn-BB84Yn},
in both the case of a large $e_{detector}$ (i.e. $\eta_n e_{detector} \gg (1-\eta_n) p_{dark}$, corresponding to Figs.~\ref{fig-decoygraph1})
and the case of $e_{detector}=0$ (corresponding to Figs.~\ref{fig-decoygraph1b}), we can see that the yields in SARG04 is about half of that in BB84;
this means that $Q_n$ in SARG04 is also about half of that in BB84.
Similarly, by comparing the bit error rates of SARG04 in \myeqnref{eqn-SARG04en} and of BB84 in \myeqnref{eqn-BB84en}, we can see that $e_n$ in SARG04 is about twice of that in BB84 for both figures;
this means that the amount of privacy amplification needed for the one-photon part of SARG04 is higher than that for BB84 (even when the mutual information between the bit and phase errors in one-photon SARG04 is taken into account).
From the key generation rate equations in \myeqnref{eqn-decoyBB84rate1} and \myeqnref{eqn-decoySARG04rate1}, the decrease in $Q_n$ and the increase in $e_n$ both reduce the key generation rate of one-photon SARG04 relative to BB84, whether or not decoy is used.
Furthermore, based on our simulations, we observe that the gap between SARG04 and BB84 decreases as $e_{detector}$ decreases.

\section{Summary and concluding remarks\label{sec:conclusion}}

We have provided lower and upper bounds on the bit error rates for SARG04 with two-way classical communications.
Both the single-photon part and the two-photon part were considered.
For the single-photon part, we have shown that SARG04 with two-way communications can tolerate a higher bit error rate than SARG04 with one-way communications.
However, it does not mean that for some smaller bit error rate, two-way SARG04 protocol has higher key generation rate than the one-way version.

The upper bounds were found by considering a general intercept-and-resend attack by Eve.
In this attack, she performs an arbitrary POVM and sends arbitrary states to Bob according to the measurement outcome.
For the one-photon case, we have shown that such generality in her attack does not offer any advantage over a simple intercept-and-resend attack where she only performs measurement and sends the measurement results to Bob.

We have also studied SARG04 with a coherent source using the decoy-state method to achieve unconditional security.
The key generation rate is significantly improved by combining the GLLP and the decoy-state ideas compared to the non-decoy protocols.
This improved key rate for SARG04 is given by~\cite{Tamaki2005}
\begin{eqnarray}
R_{\text{\tiny SARG04}} = -Q_\mu f(E_\mu) H_2 (E_\mu) + Q_1 [1-H(Z_1|X_1)] + Q_2 [1-H(Z_2|X_2)].
\end{eqnarray}
The first term is the fraction of EPR pairs spent for bit error correction, the second term is the contribution to the key rate from the single-photon states, and the third term is the contribution from the two-photon states.
In all our simulations, we found that SARG04 has a smaller key generation rate and a shorter secure distance than BB84, using the combined GLLP and decoy formulation.
Our results apply to the case where Eve performs the most general attack.
This situation is different from that in~\cite{Branciard2005}, where they assumed that Eve performs an individual attack.
We have shown that optimal mean photon number for SARG04 can be higher than that of BB84 for small misalignment errors in the detectors.
Also, we observed that the
optimal $\mu$ for SARG04 and BB84 is approximately constant for a large of distances.
This means that the key generation rates for both of SARG04 and BB84
increase linearly with the transmission efficiency $\eta$.


It is interesting to generalize our formulation of SARG04 with infinite decoys to the case of finite decoys, and to the case of using two-way classical communications with decoy.
Also, our work can be extended to generalizations of SARG04, the six-state SARG04~\cite{Tamaki2005} and the $N$-state protocol~\cite{Koashi2005}.
We leave them for future studies.



\appendix

\section{Density matrices of one- and two-photon SARG04\label{app-SARG04-densitymatrix}}

\subsection{One-photon case}
We consider the most general attack by Eve on all qubits sent by Alice.
We focus on the density matrix of one qubit, denoted as $\rho_\text{qubit}$, which is obtained by tracing out all other qubits.
Alice initially prepares $\ket{\Psi}_{A B} =  (\ket{0_z}_A \ket{\varphi_0}_B + \ket{1_z}_A \ket{\varphi_1}_B)/\sqrt{2}$ and applies a random rotation, $R^k$, on system $B$.
After Eve's attack and Bob's inverse rotation and successful filtering, the final qubit pair state for a particular pair is
\begin{eqnarray}
\rho_\text{qubit} &=&
\sum_{k=0}^3 \sum_f P((F R^{-k}E^{(f)} R^k)_B \ket{\Psi}_{AB})
\end{eqnarray}
where $P(\ket{\Phi})=\ket{\Phi}\bra{\Phi}$ is a projection operator associated with a pure state $\ket{\Phi}$, and $E^{(f)}$ is an arbitrary matrix indexed by $f$ that includes Eve's action on this qubit.
Note that $E^{(f)}$ can be dependent on Eve's action on all the other pairs.
For the moment, we consider the case that there is only one action by Eve (i.e. $f$ takes on one value).
The (unnormalized) probability of $X$, $Y$, and $Z$ errors on $\rho_\text{qubit}$ due to $E$ can be explicitly computed using \myeqnref{eqn-Bellerrors} as follows:
\begin{eqnarray}
p_I &=& \frac{1}{2} |a_{11} + a_{22}|^2 \\
p_X &=& \frac{1}{4} (|a_{12} + a_{21}|^2 + |a_{11} - a_{22}|^2) \\
p_Y &=& \frac{1}{4} ((5 a_{12} - 3 a_{21}) a_{12}^* + (-3 a_{12} + 5 a_{21}) a_{21}^* + |a_{11} - a_{22}|^2 ) \\
p_Z &=& (|a_{12}|^2 + |a_{21}|^2) + \frac{1}{2} (|a_{11} - a_{22}|^2 )
\end{eqnarray}
where $E=\begin{pmatrix} a_{11} & a_{12} \\ a_{21} & a_{22}\end{pmatrix}$.
The bit error probability is $p_{bit}=p_X+p_Y$ and the phase error probability is $p_{phase}=p_Z+p_Y$.
It can easily be shown that
\begin{eqnarray}
p_{phase}&=&\frac{3}{2} p_{bit} \\
p_Y&=&p_{bit}/2 + |a_{12} - a_{21}|^2 /2
\end{eqnarray}
Note that the above equations involve the error {\it probabilities} of the particular pair conditioned on any configurations of the events including $X$, $Y$, and $Z$ errors for all the other pairs, but not the actual error rate of a realization of the protocol.
In an actual protocol, the actual bit error rate $e_b$ is estimated and we want to relate it to the actual phase error rate $e_p$ and also to the actual $Y$ error rate $a$ (which is the counterpart of $p_Y$).
However, we may not immediately conclude that $e_p=\frac{3}{2}e_b$ and $a \geq e_b/2$ since $p_{I/X/Y/Z}$ are only the probabilities of errors conditional on the events for other pairs;
the errors of all the EPR pairs could be arbitrarily correlated.
Nevertheless, both $e_p=\frac{3}{2}e_b$ and $a \geq e_b/2$ can be justified by using Azuma's inequality~\cite{Azuma1967}.
Let $N$ be the number of EPR pairs, $L=\{ I,X,Y,Z\}$ be a label for a Pauli operator,
$n_L^{(l)}, l \in [1,N]$ be the actual number of $L$ errors on the first $l-1$ pairs, and
$p_L^{(l)}$ be the probability of having an $L$ error on the $l^{\text{th}}$ pair conditional on any configuration of the events including the actual $X/Y/Z$ error patterns on the first $l-1$ pairs. Note that we can identify $p_{L}^{(l)}$ to $p_L$.
Applying Azuma's inequality to the random variable $n_L^N - \sum_{l=1}^{N} p_{L}^{(l)}$, one can show that $\sum_{l=1}^{N} p_{L}^{(l)} \rightarrow n_L^N$ with exponentially increasing probability as $N$ increases.
Thus, after the bit error rate estimation, Alice and Bob perceive that fractions $e_b -a$, $3 e_b/2 -a$, and $a$ of EPR pairs suffer from $X$, $Z$, and $Y$ errors respectively.
They can associate this information with a density matrix to arrive at \myeqnref{eqn-single-densitymatrix}.
A similar security analysis can be found in~\cite{Boileau2005}.

\subsection{Two-photon case}

In the two-photon case, Alice prepares a three-photon system $\ket{\Psi}_{A B E1} =  (\ket{0_z}_A \ket{\varphi_0}_B\ket{\varphi_0}_{E1} + \ket{1_z}_A \ket{\varphi_1}_B\ket{\varphi_1}_{E1})/\sqrt{2}$ and applies a random rotation, $R^k \otimes R^k$, on systems $B$ and $E1$.
System $B$ is sent to Bob through Eve while system $E1$ is kept by Eve.
We analyze this case in the same as in the one-photon case.
We obtain $\rho_{qubit}$ by tracing out all other EPR pairs and system $E1$ of the pair under consideration and we arrive at
$e_p \leq x e_b + g(x), \forall x$, where $g(x)=\frac{1}{6} (3-2x+\sqrt{6-6\sqrt{2}x+4x^2})$.
In this case, we could not find any constraint on the actual fraction of $Y$ errors, $a$.
This means $e_b \geq a \geq 0$.

\section{Proof of Theorem~\ref{thm-worst-SARG04-single}\label{app-worst-SARG04-single}}

Given two initial states $(p_{X_\alpha},p_{Y_\alpha},p_{Z_\alpha})=(e_b-a_\alpha,a_\alpha,\xi e_b -a_\alpha)$ and
$(p_{X_\beta},p_{Y_\beta},p_{Z_\beta})=(e_b-a_\beta,a_\beta,\xi e_b -a_\beta)$ where $a_\beta > a_\alpha$,
we apply the same sequence of B/P steps starting with a B step to the two initial states, thus giving rise to two sequences of states (the $\alpha$ sequence and the $\beta$ sequence).
We want to show that the final state of the $\alpha$ sequence leads to a smaller key generation rate in \myeqnref{eqn-CSSkeyrate} than that of the $\beta$ sequence.
This implies that the key generation rate is an increasing function of $a$.

Starting with a pool of EPR pairs with state $(p_X,p_Y,p_Z)$, applying a B step leads to a smaller set of surviving pairs with a new state $(p_X',p_Y',p_Z')$ described by \myeqnrefs{eqn-Bstep1}-(\ref{eqn-Bstep2}).
Similarly, beginning with $(p_X,p_Y,p_Z)$, a P step leads to a new state described by \myeqnrefs{eqn-Pstep0}-(\ref{eqn-Pstep1}).

We apply a change of variables:
\begin{eqnarray}
t_Z&=& p_X+p_Y \\
t_X&=&p_Y+p_Z \\
\Delta&=&p_Z-p_Y.
\end{eqnarray}

We start with the hypothesis that in any stage of the $\alpha$ and $\beta$ sequences, $t_{Z_\beta} = t_{Z_\alpha}$, $t_{X_\beta} \leq t_{X_\alpha}$, and $\Delta_\beta \leq \Delta_\alpha$.
If this is true and if $t_{X_\alpha} \leq \frac{1}{2}$, the key generation rate, $1-H_2(t_Z)-H_2(t_X)$, at any stage of the $\alpha$ sequence is smaller and Theorem~\ref{thm-worst-SARG04-single} follows.

First, we can verify that the hypothesis is true initially by noticing that $t_{Z_\beta} = t_{Z_\alpha}=e_b$, $t_{X_\beta} = t_{X_\alpha} = \xi e_b$ and $\Delta_\beta=\xi e_b-2 a_\beta < \Delta_\alpha=\xi e_b -2 a_\alpha$.

Next, we show that given the hypothesis is true for the current stage, it is also true for the next stage when a B step is applied.
In the new variables, the new state after a B step becomes
\begin{eqnarray}
t_Z' &=& t_Z^2/p_S \\
t_X' &=& [t_X - t_X^2 + \Delta ( 1- 2 t_Z -\Delta)] / p_S \\
\label{eqn-app-delta1}
\Delta' &=& [t_X(1-2t_Z)+\Delta (1-2t_X)]/p_S \\
p_S &=& 1-2t_Z+2t_Z^2.
\end{eqnarray}
Given that $t_{X_\beta} \leq t_{X_\alpha}$ and $\Delta_\beta \leq \Delta_\alpha$, we express the state of sequence $\beta$ in terms of that of sequence $\alpha$:
\begin{eqnarray}
t_{Z_\beta}' &=& t_{Z_\alpha}' \\
t_{X_\beta}' &=& t_{X_\alpha}' - [t_{X_{\alpha-\beta}}(1-2 t_{X_\alpha}+t_{X_{\alpha-\beta}}) + \Delta_{\alpha-\beta}(1-2 t_{Z_\beta}-\Delta_\alpha-\Delta_\beta)] / p_S \\
\Delta_\beta' &=& \Delta_\alpha'- [t_{X_{\alpha-\beta}}(1-2t_{Z_\beta}-2\Delta_\beta)+ \Delta_{\alpha-\beta} (1-2t_{X_\alpha})]/p_S \\
p_S &=& 1-2t_Z+2t_Z^2,
\end{eqnarray}
where $t_{X_{\alpha-\beta}}=t_{X_\alpha}-t_{X_\beta} \geq 0$ and $\Delta_{\alpha-\beta}=\Delta_\alpha-\Delta_\beta \geq 0$.
Obviously, the hypothesis for the primed variables is true if $(1-2t_{Z}-2\Delta)\geq 0$, $t_X \leq \frac{1}{2}$, and $t_Z \leq \frac{1}{2}$ at any stage of the $\alpha$ and $\beta$ sequence.
We will show the first inequality later and
impose the last two inequalities as condition (ii) of the theorem.

We consider the new state after a P step is applied and show that the hypothesis is also true for this new state.
The new state after a P step is
\begin{eqnarray}
t_Z' &=& 3 t_Z (1-t_Z)^2 + t_Z^3 \\
t_X' &=& 3 t_X^2(1-t_X)+t_X^3 \\
\label{eqn-app-delta2}
\Delta' &=& 3 \Delta^2 ( 1-2 t_Z-\Delta)+\Delta^3.
\end{eqnarray}
It is obvious that $t_X'$ increases with $t_X$, which implies that $t_{X_\beta}' \leq t_{X_\alpha}'$.
Also, $\Delta'$ increases with $\Delta$ provided that $( 1-2 t_Z-\Delta)\geq 0$ and $\Delta \geq0$, which implies that $\Delta_\beta' \leq \Delta_\alpha'$.
The first inequality is satisfied if $(1-2t_{Z}-2\Delta)\geq 0$, which will be shown later.
We first show that $\Delta \geq 0$.

\begin{claim}
After the initial B step, or after any B/P step that follows, $\Delta \geq 0$ holds.
\end{claim}
\begin{proof}
Before the initial B step is applied, we have
\begin{eqnarray}
\Delta &=& \xi e_b -2a \\
&\geq& \xi e_b -2 e_b \\
&\geq& -e_b,
\end{eqnarray}
where the last inequality is due to $\xi \geq 1$.
After the initial B step, from \myeqnref{eqn-app-delta1}, we have $\Delta' \geq 0$ if the following condition is satisfied:
\begin{eqnarray}
\Delta &\geq& -\frac{\xi e_b (1-2e_b)}{1-2\xi e_b}.
\end{eqnarray}
Since the right-hand side is smaller than $-e_b$, this condition is satisfied after the first B step, which means that $\Delta' \geq 0$ after the first B step.
Furthermore, from \myeqnref{eqn-app-delta1} and \myeqnref{eqn-app-delta2}, we conclude that $\Delta' \geq 0$ after any B step or P step following the initial B step.
\end{proof}

\begin{claim}
$1-2t_Z-2\Delta\geq 0$ always holds if $e_b \leq \frac{1+4a}{2(1+\xi)}$ (which is condition (i) of the Theorem).
\end{claim}
\begin{proof}
Before the initial B step, we can easily see
\begin{eqnarray}
1-2t_Z-2\Delta &=& 1-2(1+\xi)e_b+4a \geq 0
\end{eqnarray}
because $e_b \leq \frac{1+4a}{2(1+\xi)}$.
After a B step,
\begin{eqnarray}
1-2t_Z'-2\Delta' &=& 1- [2 t_Z^2 + 2 t_X (1-2t_Z)+2\Delta(1-2t_X)]/p_S \\
&=&(1-2t_Z-2\Delta)(1-2t_X)/p_S,
\end{eqnarray}
which is non-negative when $1-2t_Z-2\Delta\geq 0$.

After a P step,
\begin{eqnarray}
1-2t_Z' &=& (1-2t_Z)^3
\end{eqnarray}
so
\begin{eqnarray}
1-2t_Z'-2\Delta' &=& (1-2t_Z)^3-6\Delta^2(1-2t_Z)+4 \Delta^3 \\
&=& (1-2t_Z-2\Delta) [ (1-2t_Z)^2 + 2 \Delta (1-2t_Z-\Delta)].
\end{eqnarray}
which is non-negative when $1-2t_Z-2\Delta\geq 0$ and $\Delta \geq 0$.
\end{proof}

\section{Proof of Theorem~\ref{thm-SARG04-generalattack1} and Theorem~\ref{thm-SARG04-generalattack2}\label{app-SARG04-generalattack}}

In this appendix, we will prove that a general POVM attack by Eve induces a bit error rate of at least $\frac{1}{3}$ for the single-photon case.
To do this, we will first consider a special case of this attack where Eve always sends only SARG04 states to Bob.
Then building on the proof of this special case, we will show that the minimum bit error rate is $\frac{1}{3}$ for the general POVM attack where Eve sends arbitrary states to Bob.
At last, we will generalize the proof to the case of two photons, showing that it is possible to derive the minimum bit error rate even for this case.

Before we begin, we note that $R^4 = I$.
This allows us to adopt the following notation:
\begin{eqnarray}
\label{eqn-SARG04-generalattack-1}
\ket{\varphi_{m+k}}&=&R^{-m} \ket{\varphi_k}, \: \forall m,k \in \field{Z}
\end{eqnarray}
where the subscripts of the SARG04 states are taken in {\em module} $4$.


\subsection{Eve sending SARG04 states}

\begin{figure}
\centering
\includegraphics[width=\figlength]{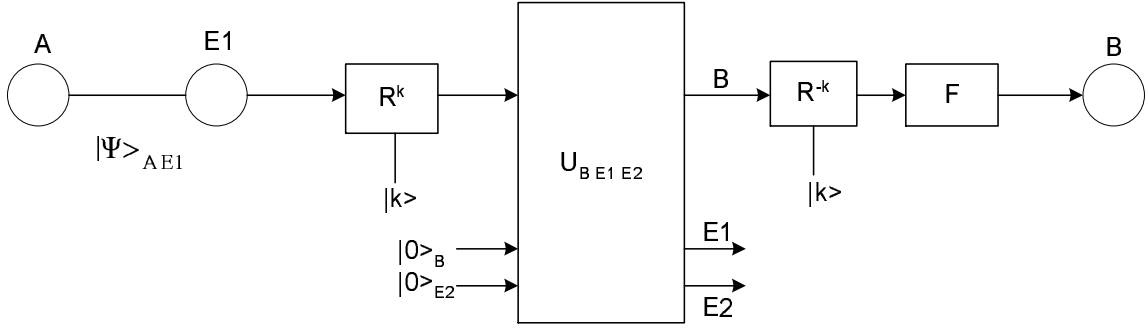}
\caption{\label{fig-attack-1}A POVM attack by Eve realized by $U_{B E_1 E_2}$.}
\end{figure}

A block diagram showing an attack by Eve is depicted in Fig.~\ref{fig-attack-1}.
First, Alice prepares a bipartite entangled state $\ket{\Psi}_{A E_1} =  \ket{0_z}_A \ket{\varphi_0}_{E_1} + \ket{1_z}_A \ket{\varphi_1}_{E_1}$.
After randomly applying a rotation $R^k$, she sends the $E_1$ qubit to Eve, who will then perform a POVM $\{W_m^\dagger W_m\}$ on $E_1$, which is realized by an unitary operator $U_{B E_1 E_2}$.
When the measurement result is $m$, Eve sends a state $\ket{\varphi_m}_B$ to Bob.
We will obtain the density matrix of Alice and Bob $\rho_{AB}$ and minimize the bit error rate $[Tr(\rho_{AB})]^{-1}[\bra{\Psi^+}\rho_{AB}\ket{\Psi^+}+\bra{\Psi^-}\rho_{AB}\ket{\Psi^-}]$ over $W_m$'s.

The input state transforms as follows:
\begin{eqnarray}
&&\sum_k ( \mathbf{I}_A \otimes R^k_{E_1}) \ket{\Psi}_{A E_1} \ket{k}_K \\
&\xrightarrow{U} & \sum_k \sum_{m=0}^{3} ( \mathbf{I}_A \otimes (W_m R^k)_{E_1}) \ket{\Psi}_{A E_1} \ket{m}_{E_2} \ket{\varphi_m}_B \ket{k}_K \\
&\xrightarrow{R^{-k} F} & \sum_k \sum_{m=0}^{3} \big[\ket{0_z}_A ( W_m R^k \ket{\varphi_0}_{E_1}) + \ket{1_z}_A (W_m R^k \ket{\varphi_1}_{E_1})\big] \otimes \\
&&\ket{m}_{E_2} (F R^{-k} \ket{\varphi_m}_B) \ket{k}_K \nonumber
\end{eqnarray}
We then trace out systems $E_1$, $E_2$, and $K$ to get the final density matrix between Alice and Bob:
\begin{eqnarray}
\rho_{AB}
&=& \sum_{k=0}^{3} \sum_{m=0}^{3}
\Big( a^{00}_{mk}\ket{0_z}_A\bra{0_z}
+a^{01}_{mk}\ket{0_z}_A\bra{1_z}
+a^{10}_{mk}\ket{1_z}_A\bra{0_z}
+a^{11}_{mk}\ket{1_z}_A\bra{1_z} \Big) \\
&& \otimes F \ket{\varphi_{m+k}}_B \bra{\varphi_{m+k}} F^\dagger
\end{eqnarray}
where
\begin{eqnarray}
a^{00}_{mk} &=& | \bra{0_z}W_m\ket{\varphi_{-k}} | ^2 + | \bra{1_z}W_m\ket{\varphi_{-k}} | ^2 \\
a^{01}_{mk} &=& \bra{0_z}W_m\ket{\varphi_{-k}} \bra{0_z}W_m\ket{\varphi_{1-k}}^* + \bra{1_z}W_m\ket{\varphi_{-k}} \bra{1_z}W_m\ket{\varphi_{1-k}}^* \\
a^{10}_{mk} &=& (a^{01}_{mk} )^* \\
a^{11}_{mk} &=& | \bra{0_z}W_m\ket{\varphi_{1-k}} | ^2 + | \bra{1_z}W_m\ket{\varphi_{1-k}} | ^2
\end{eqnarray}
Here, we have used the notation in \myeqnref{eqn-SARG04-generalattack-1}.
Note that $\rho_{AB}$ is a separable density matrix as we have explicitly constructed it to be, and because of that, no entanglement exists and thus no secure key can be distilled.
We can compute the unnormalized bit error rate $p_X+p_Y$ as
\begin{eqnarray}
p_X+p_Y
&=& \phantom{}_{AB}\bra{0_z 1_z} \rho_{AB} \ket{0_z 1_z}_{AB} + \phantom{}_{AB}\bra{1_z 0_z} \rho_{AB} \ket{1_z 0_z}_{AB} \\
&=& \sum_{k+m=0} a^{11}_{mk}\frac{1}{4} + \sum_{k+m=1} a^{00}_{mk} \frac{1}{4} \nonumber \\
&&+ \sum_{k+m=2} \big( a^{00}_{mk} \frac{1}{2} + a^{11}_{mk}\frac{1}{4} \big) \\
&&+ \sum_{k+m=3} \big( a^{11}_{mk} \frac{1}{2} + a^{00}_{mk}\frac{1}{4} \big) \nonumber \\
&=& \sum_{m=0}^{3} \sum_{j=0}^{1} \bra{j_z}W_m L_m W_m^\dagger \ket{j_z},
\end{eqnarray}
where
\begin{eqnarray}
L_m &=& \frac{1}{2} \ket{\varphi_{1+m}}\bra{\varphi_{1+m}}
 + \ket{\varphi_{2+m}}\bra{\varphi_{2+m}}
 +\frac{1}{2} \ket{\varphi_{3+m}}\bra{\varphi_{3+m}}.
\end{eqnarray}
Since $W_m$ is some $2 \times 2$ matrix (not necessary Hermitian), the problem of finding $W_m$ is broken into finding two independent $1 \times 2$ vectors $\bra{0_z}W_m$ and $\bra{1_z}W_m$.

In order to normalize the bit error rate, we find
\begin{eqnarray}
Tr(\rho_{AB}) &=&
\sum_{i,j \in \{0,1\}} \phantom{}_{AB}\bra{i_z j_z} \rho_{AB} \ket{i_z j_z}_{AB} \\
&=& \sum_{m=0}^{3} \sum_{j=0}^{1} \bra{j_z}W_m B_m W_m^\dagger \ket{j_z},
\end{eqnarray}
where
\begin{eqnarray}
B_m &=&
\frac{1}{2} \ket{\varphi_{0+m}}\bra{\varphi_{0+m}}
 +\ket{\varphi_{1+m}}\bra{\varphi_{1+m}} \nonumber \\
&&
 + \frac{3}{2} \ket{\varphi_{2+m}}\bra{\varphi_{2+m}}
 +\ket{\varphi_{3+m}}\bra{\varphi_{3+m}}.
\end{eqnarray}
Therefore, the normalized bit error rate is
\begin{eqnarray}
e_b &=& \frac{
\sum_{m=0}^{3} \sum_{j=0}^{1} \bra{j_z}W_m L_m W_m^\dagger \ket{j_z}
}{
\sum_{m=0}^{3} \sum_{j=0}^{1} \bra{j_z}W_m B_m W_m^\dagger \ket{j_z}
}
\end{eqnarray}
We want to minimize $e_b$ over the eight independent $1 \times 2$ vectors $\bra{j_z}W_m$.
At least one of the eight must be non-zero, otherwise all $W_m$ would be zero and there would be no qubits sent to Bob.
Since $e_b$ is not a sum of eight independent ratios, i.e.
\begin{eqnarray}
e_b &\neq& \sum_{m=0}^{3} \sum_{j=0}^{1}
\frac{
 \bra{j_z}W_m L_m W_m^\dagger \ket{j_z}
}{
 \bra{j_z}W_m B_m W_m^\dagger \ket{j_z}
},
\end{eqnarray}
it may appear at first sight that the minimization of $e_b$ is not trivial.
However, it turns out that we can minimize each ratio independently and set $e_b$ to be the smallest ratio by assigning zeros to the other seven vectors.
We show this by the following claim:
\begin{claim}
\label{claim-SARG04-generalattack-ratio}
Given two ratios, $\frac{a_1}{a_2}$ and $\frac{b_1}{b_2}$,
if $\frac{a_1}{a_2} \leq \frac{b_1}{b_2}$,
then $\frac{a_1}{a_2} \leq \frac{a_1+b_1}{a_2+b_2}$.
\end{claim}
Therefore, we consider separately minimizing each ratio, which can be written as
\begin{eqnarray}
\frac{
\langle c_{jm}| B_m^{-\frac{1}{2}} L_m B_m^{-\frac{1}{2}} |c_{jm}\rangle
}{
\langle c_{jm}|c_{jm}\rangle
}
\end{eqnarray}
where $\langle c_{jm}|=\bra{j_z}W_m B_m^{\frac{1}{2}}$ is a $1 \times 2$ vector.
The minimizing $c_{jm}$ is the eigenvector of $B_m^{-\frac{1}{2}} L_m B_m^{-\frac{1}{2}}$ corresponding to the minimum eigenvalue.
The two eigenvalues are $0.6$ and $\frac{1}{3}$ for all $m$.
Thus, the minimum $e_b$ is $\frac{1}{3}$.
A POVM $\{W_m^\dagger W_m\}$ that is compatible with these eigenvectors is $W_m^\dagger W_m=\ket{\varphi_m}\bra{\varphi_m}/2, m=0,\ldots,3$, which is the trivial intercept-and-resend attack.

\subsection{Eve sending arbitrary states}

Now, instead of sending the four SARG04 states $\ket{\varphi_i},i=0,\ldots,3$, we assume Eve sends
any number, $G$, of arbitrary states.
We label these states as $\ket{\sigma_0^g}, \: g=0,\ldots,G-1$.
For the sake of making the analysis of this case parallel to that of the previous case of sending SARG04 states,
we associate three extra states (with certain symmetry) to each arbitrary state and we label all states as follows:
\begin{eqnarray}
\ket{\sigma_i^g}, \: i=0,\ldots,3, \: g=0,\ldots,G-1.
\end{eqnarray}
We can view the states as divided into sets of four with a total of $G$ sets.
The $i=0$ states are the original arbitrary states and are called the representative states of its set;
the $i=1,2,3$ states are the extra states introduced.
The POVM elements $\{{W_i^g}^\dagger W_i^g\}$ corresponding to the states are also indexed in the same way.
Along the same lines as the SARG04 states, we define the extra states to have a rotational symmetry that satisfies $\ket{\sigma_{m+k}^g}=R^{-k} \ket{\sigma_m^g}, \forall g$.
This symmetry requirement makes the analysis much easier since it resembles the analysis for the case of sending SARG04 states.
Note that the introduction of the three extra states in each set does not lose any generality, since
if the extra states are not needed in the minimization of the bit error rate, their corresponding POVM elements will eventually be found to be zeros.

The analysis of this case basically goes as before by replacing $\ket{\varphi_i}$ with $\ket{\sigma_i^g}$.
The final normalized bit error rate is
\begin{eqnarray}
e_b &=& \frac{
\sum_{g=0}^{G-1} \sum_{m=0}^{3} \sum_{j=0}^{1} \bra{j_z}W_m^g L_m^g {W_m^g}^\dagger \ket{j_z}
}{
\sum_{g=0}^{G-1} \sum_{m=0}^{3} \sum_{j=0}^{1} \bra{j_z}W_m^g B_m^g {W_m^g}^\dagger \ket{j_z}
},
\end{eqnarray}
which has the same form as before but with different $L_m^g$'s and $B_m^g$'s.
As before, both of them are weighted sums of the outer products of the SARG04 states, $\sum_{i=1}^4 \kappa_{im} \ket{\varphi_{i+m}}\bra{\varphi_{i+m}}$.  ($\kappa_{im}$'s for $B_m^g$ and $L_m^g$ are different.)
The difference is that now $\kappa_{im}$'s are no longer constant, but dependent on the representative state of each set sent by Eve, $\ket{\sigma_0^g}$.
Thus, $W_m^g$ is also a function of this state.
Since Claim~\ref{claim-SARG04-generalattack-ratio} says that we can minimize each term of $e_b$ separately and since $\ket{\sigma_0^g}$ is arbitrary anyway, we only need to focus on $L_0^0$ and $B_0^0$ and minimize the eigenvalues of ${(B_0^0)}^{-\frac{1}{2}} L_0^0 {(B_0^0)}^{-\frac{1}{2}}$ (which correspond to the bit error rate).
The two eigenvalues are
\begin{eqnarray}
\frac{2-c}{4-c}
&\text{and}&
\frac{2+c}{4+c}.
\end{eqnarray}
where 
$\ket{\sigma_0^0} \triangleq \sigma_{00} \ket{0_z} + \sigma_{01} \ket{1_z}$,
$c=\frac{| \sigma_{00}^2 + \sigma_{01}^2 |}{| \sigma_{00} |^2 + | \sigma_{01} |^2} \leq 1$.
The minimum of the first eigenvalue is $\frac{1}{3}$ at $c=1$ and the second eigenvalue is in $[0.5,0.6]$.
Therefore, we conclude that, for the one-photon SARG04 case, the minimum bit error rate caused by Eve using a general POVM intercept-and-resend attack with arbitrary states sent is $\frac{1}{3}$.
Note that $c=1$ corresponds to the phase difference between $\sigma_{00}$ and $\sigma_{01}$ being $0$ or $\pi$, under our specific choice of the SARG04 states.
Also, the bit error rate of $\frac{1}{3}$ can be achieved with any assignment of $\sigma_{00}$ and $\sigma_{01}$ (of course, different assignments of them give rise to different POVM elements), as long as they are in phase or completely out of phase.

\subsection{Two-photon case}
We can extend this proof to the two-photon SARG04 case easily.
The initial state becomes $\ket{\Psi}_{A E_1} =  \ket{0_z}_A \ket{\varphi_0\varphi_0}_{E_1} + \ket{1_z}_A \ket{\varphi_1\varphi_1}_{E_1}$, as Alice emits two photons to Eve.
Alice applies rotation $R^k \otimes R^k$ to the two-qubit system $E_1$ before it is sent to Eve.
Eve then performs a POVM on $E_1$ and, based on the measurement outcome, sends system $B$ to Bob as before.
The analysis for this case is the same as the one-photon case, with the change of $E_1$ being a two-qubit system.
Because of this change,
the matrices $W_m^g$, $L_m^g$, and $B_m^g$ in the analysis are subsequently changed to have dimension $4 \times 4$.
Both $L_m^g$ and $B_m^g$ are enlarged by replacing every tensor product of the form $\ket{\varphi_{m}}\bra{\varphi_{m}}$ by $\ket{\varphi_{m}\varphi_{m}}\bra{\varphi_{m}\varphi_{m}}$, with no change to the corresponding coefficients.
We can carry the same analysis as the single-photon case and arrive at the eigenvalues of $(B_0^0)^{-\frac{1}{2}} L_0 (B_0^0)^{-\frac{1}{2}}$ to determine the bit error rate\footnote{Actually, the pseudo inverse of $B_0^0$ is used since $B_0^0$ (and $L_0^0$) has rank 3.  The analysis is not affected since the nullspaces of $B_0^0$ and $L_0^0$ are the same.}.
Because of the increased dimension in this case, we could not directly solve for the eigenvalues in terms of $\ket{\sigma_0^g}$.
Instead, we parameterize the eigenvalues with two parameters, $\theta_z$, and $\theta_y$, and plot the eigenvalues against these two parameters.
These two parameters come from the fact that any state can be written as a rotation about the $z$-axis on $\ket{\varphi_0}$ (which is not equal to $\ket{0_z}$ or $\ket{1_z}$) followed a rotation about the $y$-axis, i.e. $\ket{\sigma_0^0} = R_y(\theta_y) R_z(\theta_z) \ket{\varphi_0}$.
Using this definition for $\ket{\sigma_0^0}$, we found, from plots of the eigenvalues as functions of $\theta_y$ and $\theta_z$, that the eigenvalues are not dependent on $\theta_y$ and reach minimum when $\theta_z=0,\pi$.
The minimum eigenvalue (and thus the minimum bit error rate) is $\frac{3-\sqrt{2}}{7} \approx 22.65\%$.
A POVM that gives rise to this minimum bit error rate is
\begin{eqnarray}
W_m^\dagger W_m &=& P(\lambda_+ \ket{\varphi_m} \ket{\varphi_m} + \lambda_- \ket{\varphi_{m+2}} \ket{\varphi_{m+2}})
\: , m=0,\ldots,3 \\
W_{\text{vac}}^\dagger W_{\text{vac}} &=&
P(\ket{\varphi_0} \ket{\varphi_2} - \ket{\varphi_2} \ket{\varphi_0})/2 \\
&=&P(\ket{\varphi_3} \ket{\varphi_1} - \ket{\varphi_1} \ket{\varphi_3})/2
\end{eqnarray}
where $\lambda_{\pm}=(\pm2+\sqrt{2})/4$ and $P(\ket{\Phi})=\ket{\Phi}\bra{\Phi}$ is a projection operator associated with a pure state $\ket{\Phi}$.
Eve sends $\ket{\varphi_m}$ to Bob when the measurement outcome is $m \in [0,3]$.
Note that $W_{\text{vac}}^\dagger W_{\text{vac}}$ never occurs, since the four states sent by Alice, $\ket{\varphi_m} \ket{\varphi_m}, m \in [0,3]$, are orthogonal to the state $W_{\text{vac}}^\dagger W_{\text{vac}}$ projects onto.

\bibliographystyle{apsrev.bst}
\bibliography{supp/paperdb}

\end{document}